\documentstyle[aps,epsfig,bbm]{revtex}
\begin{document}

\draft
\title{Coherent backscattering of light by atoms in the saturated regime}
\author{T. Wellens$^{1,2}$, B. Gremaud$^2$, D. Delande$^2$, and
C. Miniatura$^1$}
\address{$^1$Institut Non Lin{\'e}aire de Nice, 1361 route 
des Lucioles, F-06560 Valbonne\\
$^2$ Laboratoire Kastler Brossel, Universit{\'e} Pierre et Marie Curie,
4 Place Jussieu, F-75005 Paris}

\date{\today}

\twocolumn
\maketitle

\begin{abstract}
We present the first calculation of coherent backscattering with inelastic
scattering by saturated atoms. We consider the scattering of a 
quasi-monochromatic laser pulse by
two distant
atoms in free space. By restricting ourselves to scattering of two photons,
we employ a perturbative approach,
valid up to second order in the incident laser
intensity. The  backscattering enhancement factor is found to be smaller than
two (after excluding single scattering), indicating a loss of coherence
between the doubly scattered light emitted by both atoms. Since the
undetected photon carries information about the path of
the detected photon, the coherence loss can be explained by a which-path
argument, in analogy with a double-slit experiment.
\end{abstract}


\section{Introduction}

Weak localization of light in random media was demonstrated for the first time
in the eighties \cite{kuga84,albada85,wolf85}.
Here, constructive interference between two waves
which interact with the same
particles, but in reversed order,
enhances, in average, 
scattering in the direction opposite to the
incident light. For systems obeying the reciprocity symmetry \cite{vanT97},
the {\em backscattering enhancement factor}, i.e. the light intensity
detected in exact backscattering direction divided by the background
intensity, is exactly two, provided single scattering can be removed.
For spherically symmetric scatterers, the latter is achieved in the helicity
preserving polarization channel.

Similar interference effects between multiply scattered waves also affect
the properties of
transport through disordered media. If the mean free path
can be sufficiently reduced, the transport is even expected to come to
a complete standstill \cite{chabanov00}. In experiments on
strong localization of light \cite{wiersma97}, however, the role
of absorption is discussed controversially \cite{scheffold99,wiersma99}.

One may wonder whether a medium consisting of individual atoms would
constitute a good candidate for strong localization. 
In contrast to the classical scenario (Maxwell's equations
in a medium with random dielectric constant), the quantum-mechanical
atom-photon interaction exhibits some characteristic features,
which may affect the coherence between multiply scattered
waves. 
Firstly, the resonance may be extremely sharp, corresponding to a
very narrow linewidth $\Gamma$ of the excited state. 
On the one hand, this leads to a large atom-photon
scattering cross section and slow diffusion of light \cite{labeyrie03} - 
properties in favor of localization.
On the other hand, it implies that the
atoms have to be cooled to very low temperatures. Only if the
Doppler shift induced by a moving atom is much smaller than
$\Gamma$, the interference between 
two counterpropagating waves is preserved \cite{wil03}.
Typically, this regime is reached at about a
few $\rm mK$, which is, however, still
high enough to neglect the thermal de-Broglie wavelength of the atoms, i.e. to
treat their external motion classically.
Furthermore, atoms usually have an internal quantum structure, which may have
a strong impact on coherent backscattering 
\cite{cord1,cord2,labeyrie99,jonkheere00}.
If necessary, this can
be circumvented by using atoms with a nondegenerate ground state ($J=0$)
\cite{bidel02}.

Another property of the atom-light interaction, whose impact on coherent
backscattering has so far remained almost unexplored,
is the strongly nonlinear response of
an atom to incoming
radiation. A single photon being sufficient to bring the atom to the
excited state,
where it rests for a quite long time $\Gamma^{-1}$ without being able to 
scatter other photons, a saturation of the atomic medium can
be induced already with rather moderate laser intensities. 
Not only the atom-photon cross section, but also the
spectrum of the light is affected by saturation.
With increasing saturation, it becomes more and more probable that
an atom scatters inelastically, i.e. that it emits photons 
at a frequency different from the one of the incident laser.
As we will show in this paper, this implies a loss of coherence
between two reversed scattering paths.
Similarly, a recent experiment showed
coherent backscattering  by a cloud of cold Strontium atoms
to be reduced when increasing the saturation induced by
the probe laser \cite{Chan03}.

In order to expose the physical mechanism responsible for the loss
of coherence as clearly as possible, we will consider
in this paper two two-level atoms in free space - 
the simplest system exhibiting coherent backscattering. Effects which
arise in the presence of a larger number of atoms,
such as the nonlinear index of refraction of an atomic medium,
will be relegated to future publications.
Furthermore, we assume that the distance between the two atoms 
is much larger than the optical wavelength, such that we can neglect 
recurrent scattering (corresponding to a dilute medium in the case of
many atoms). To calculate the photodetection signal of the light
emitted by the two atoms, we use scattering theory.
Generally, the higher the intensity of the incoming light, the more
photons are scattered inelastically. In the present paper,
we restrict ourselves to two-photon scattering. 
Thereby, we employ a perturbative approach, valid up to second order in the
incident intensity. 

The paper is organized as follows. In Sec.~\ref{single}, we 
summarize known results about the scattering of two photons by a single atom.
After introducing
the scattering operator in Sec.~\ref{scatter1},
we obtain the corresponding photodetection signal in Sec.~\ref{spower},
thereby recovering the resonance fluorescence spectrum  
in second order of the intensity.
In Sec.~\ref{double}, we add a second atom to our model. We
proceed in a similar way as in Sec.~\ref{single}, using the results of the
single-atom case as a building block of the two-atom solution.
After deriving 
the scattering operator in Sec.~\ref{scatter2}, we
calculate the photodetection signal in Sec.~\ref{backscattering}.
In contrast to the single-atom case, the latter contains interference
between the light emitted by the two atoms, enhancing the
detection signal in the backscattering direction. In this way, we obtain the
main result of the present paper, the backscattering enhancement factor,
which is found to be smaller than two, due to inelastic scattering.
This fact is interpreted in 
Sec.~\ref{interpret} as a loss of coherence between the light scattered by 
both atoms in opposite order. Regarding the undetected photon as a path
detector for the
detected photon, we can explain the loss of coherence by an analogy to
the double-slit experiment, where the interference pattern is washed out
if we try to observe which slit the particle has passed through.
Finally, Sec.~\ref{concl} concludes the paper.

\section{Single atom}
\label{single}

Let us start with discussing the scattering of two photons by a single atom.
This is useful since we will assume later that the second atom
is far away from the first one. The two-atom 
scattering process can then be viewed at as
a succession of two single-atom scattering processes. 

\subsection{Approximations and Hamiltonian}

We assume a two-level atom located at a fixed position ${\bf r}$.
As already mentioned
above, neglecting the external atomic motion is justified at
very low temperatures, where the Doppler shift induced by the atomic
motion is small enough. Also the recoil effect, i.e. the change of the
atomic velocity when scattering a photon, can be neglected - provided that
the number of scattering events is not too large. On the other hand, the
temperature should still be high enough such that
the external atomic motion need
not to be treated quantum mechanically. Furthermore, let us stress that
we consider an {\em undegenerate} atomic ground state ($J=0$). This is
important since coherent backscattering may be severely affected by
degeneracy \cite{cord1,cord2,labeyrie99,jonkheere00}.
The excited state is then threefold degenerate ($J=1$). Which one
of the three excited states is populated depends on the polarization
of the absorbed photon.  

With the approximations mentioned above,
our Hamiltonian reads as follows: 
\begin{equation}
H=H_0+V \label{hamil},
\end{equation}
where
\begin{eqnarray}
H_0 & = & \tilde{\omega}_{\rm at}
\sigma^\dagger\sigma+\sum_{{\bf k},s}\omega_{\bf k}
a_{{\bf k}s}^\dagger a_{{\bf k}s},\label{h0}\\
V & = & \sum_{{\bf k},s}\left(ig e^{i{\bf k}{\bf r}}(\sigma^\dagger\cdot
\epsilon_{{\bf k}s}) 
a_{{\bf k}s}-ig e^{-i{\bf k}{\bf r}} 
(\sigma\cdot\epsilon^*_{{\bf k}s}) a^\dagger_{{\bf k}s}\right),\label{v}
\end{eqnarray}
denote the free evolution, and the interaction, respectively (in units
where $\hbar=1$).
Here, the operators $\sigma^\dagger$ and $\sigma$ 
describe transitions
between the atomic ground and excited states,
with energy difference $\tilde{\omega}_{\rm at}$ (in the case of an
isolated atom),
whereas
$a_{{\bf k}s}^\dagger$ and $a_{{\bf k}s}$ create and
annihilate a photon in mode $\bf k$ (a plane wave with wavevector $\bf k$)
and polarization $\epsilon_{{\bf k}s}$ (perpendicular to $\bf k$).
The coupling constant 
\begin{equation}
g=d\left(\frac{\omega_{\bf k}}{2\epsilon_0 L^3}\right)^{1/2}\simeq
d\left(\frac{\omega_{\rm at}}{2\epsilon_0 L^3}\right)^{1/2},
\label{ek}
\end{equation}
with $L^3$ the quantization volume (which will finally drop out 
of the equations, when taking the limit $L\to\infty$)
and $d$ the magnitude of the
atomic dipole, determines the strength of the atom-field coupling.

In Eq.~(\ref{v}), we have employed the so-called
\lq rotating wave approximation\rq: a transition
from one of the excited states to the ground state 
is only possible by emitting a photon,
and vice versa by absorption.
This is justified since we will restrict
ourselves to near-resonant processes,
where only photons with frequencies close
to the atomic resonance are important
(i.e. $|\omega_{\bf k}-\omega_{\rm at}|\ll\omega_{\rm at}$).
For the same reason,
we may assume a constant value of $g$ in Eq.~(\ref{ek}), i.e.
neglect its dependence on $\omega_{\bf k}$.

Due to the coupling to the electromagnetic vacuum,
the state $|e\rangle$ is unstable: after an average lifetime given by
\begin{equation}
\Gamma=\frac{d^2\omega_{\rm at}^3}{3\pi\epsilon_0}=
\frac{2g^2 \omega_{\rm at}^2 L^3}{3\pi},
\label{gamma}
\end{equation}
an excited atom decays into the ground state, through spontaneous emission
of a photon. This gives rise to an effective,
complex atomic resonance frequency
\begin{equation}
\omega_0=\omega_{\rm at}-i\frac{\Gamma}{2},
\end{equation}
where also the real part $\omega_{\rm at}$ is shifted, as compared to the
isolated atom, Eq.~(\ref{h0}).

\subsection{Scattering matrix}
\label{scatter1}

In the following, we make use of scattering theory in order to
calculate the properties of the light emitted by the atoms.
Here, the object of interest is the scattering operator $S$, which
connects the initial and final photon states $|i\rangle$ and $|f\rangle$: 
\begin{equation}
|f\rangle=S|i\rangle.
\end{equation}
The initial and final state of the atom is always the ground state
$|g\rangle$, which we do not explicitly write in the following. 
Furthermore, we will restrict ourselves to the scattering of two photons,
thereby employing a perturbative approach, valid up to second order in the
incident intensity. 

Since, as we will see below, the two photons may be scattered independently
from each other, we consider first the scattering of a 
single photon.
In order to distinguish between the scattered and non-scattered 
part of the photon wavepacket, the transition
operator $T_1$ is introduced as follows:
\begin{equation}
S_1={\mathbbm 1}-2\pi i\delta(\omega_f-\omega_i) T_1\label{ss1},
\end{equation}
where the $\delta$-function implies conservation of the
photon's frequency (which follows from energy conservation,
since the state of the atom is the same before and after scattering).
For one-photon states, its matrix elements read \cite{CCT}: 
\begin{equation}
\langle {\bf k}_f\epsilon_f|T_1|{\bf k}_i\epsilon_i\rangle=
\frac{g^2}{\omega_i-\omega_0}
(\epsilon_i\cdot\epsilon_f^*)e^{i({\bf k}_i-{\bf k}_f)\cdot {\bf r}}.\label{t1}
\end{equation} 

The situation changes when considering a second photon.
It is convenient to write the matrix elements in the following
form:\footnote{Eq.~(\ref{ss2}) is valid only if
${\bf k}_1\epsilon_1\neq {\bf k}_2\epsilon_2$ and
${\bf k}_3\epsilon_3\neq {\bf k}_4\epsilon_4$. We will not
consider double occupancy of modes in the following, since it can be neglected
in the continuous limit of infinite mode density. In other words: two photons
are never exactly in the same mode, although they
may be infinitesimally close to each other.}
\begin{eqnarray}
\langle {\bf k_3}\epsilon_3,{\bf k_4}\epsilon_4|S_2|{\bf k_1}\epsilon_1,{\bf k_2}\epsilon_2\rangle & = &\nonumber\\
& & \!\!\!\!\!\!\!\!\!\!\!\!\!\!\!\!\!\!\!\!\!\!\!\!\!\!\!\!\!\!\!\!
\!\!\!\!\!\!\!\!\!\!\!\!\!\!\!\!\!\!\!\!\!\!\!\!\!\!\!\!\!\!\!\!
\langle {\bf k_3}\epsilon_3|S_1|{\bf k_1}\epsilon_1\rangle\langle {\bf k_4}\epsilon_4|S_1|{\bf k_2}\epsilon_2\rangle
\nonumber\\
& & \!\!\!\!\!\!\!\!\!\!\!\!\!\!\!\!\!\!\!\!\!\!\!\!\!\!\!\!\!\!\!\! +
\langle {\bf k_3}\epsilon_3|S_1|{\bf k_2}\epsilon_2\rangle\langle
{\bf k_4}\epsilon_4|S_1|{\bf k_1}\epsilon_1\rangle\nonumber\\
& & \!\!\!\!\!\!\!\!\!\!\!\!\!\!\!\!\!\!\!\!\!\!\!\!\!\!\!\!\!\!\!\!
\!\!\!\!\!\!\!\!\!\!\!\!\!\!\!\!\!\!\!\!\!\!\!\!\!\!\!\!\!\!\!\!
+ \langle {\bf k_3}\epsilon_3,{\bf k_4}\epsilon_4|T_2|
{\bf k_1}\epsilon_1,{\bf k_2}\epsilon_2\rangle.
\label{ss2}
\end{eqnarray}
Here, the first two terms scatter the two photons independently
from each other.
(There are two terms since the photons are indistinguishable: the
final photon $|{\bf k_3}\epsilon_3\rangle$, for
example, may correspond either to the initial photon
$|{\bf k_1}\epsilon_1\rangle$ or $|{\bf k_2}\epsilon_2\rangle$.) 
Since, however, the atom cannot interact with the second photon while
it is excited by the first one, the photons are in fact not completely
independent. This gives rise to the second term \cite{Dal83}: 
\begin{eqnarray}
\langle {\bf k_3}\epsilon_3,{\bf k_4}\epsilon_4|T_2|
{\bf k_1}\epsilon_1,{\bf k_2}\epsilon_2\rangle
& = & \nonumber\\
& & \!\!\!\!\!\!\!\!\!\!\!\!\!\!\!\!\!\!\!\!\!\!\!\!\!\!\!\!\!\!
\!\!\!\!\!\!\!\!\!\!\!\!\!\!\!\!\!\!\!\!\!\!\!\!\!\!\!\!\!\!\!\!\!
\!\!\!\!\!\!\!\!\!\!\!
2\pi i\frac{g^4 \delta(\omega_1+\omega_2-\omega_3-\omega_4)}
{(\omega_1-\omega_0)(\omega_2-\omega_0)}
\left(\frac{1}{\omega_3-\omega_0}+\frac{1}{\omega_4-\omega_0}\right)
\nonumber\\
& & \!\!\!\!\!\!\!\!\!\!\!\!\!\!\!\!\!\!\!\!\!\!\!\!\!\!\!\!\!\!
\!\!\!\!\!\!\!\!\!\!\!\!\!\!\!\!\!\!\!\!\!\!\!\!\!\!\!\!\!\!\!\!\!\!\!\!\!
\!\!\!\!\!\!
[(\epsilon_1\cdot\epsilon_3^*)(\epsilon_2\cdot\epsilon_4^*)+
(\epsilon_2\cdot\epsilon_3^*)(\epsilon_1\cdot\epsilon_4^*)]
e^{i({\bf k}_1+{\bf k}_2-
{\bf k}_3-{\bf k}_4)\cdot{\bf r}}.
\label{t2}
\end{eqnarray}
Although their sum is conserved, the individual frequencies of both photons
may be changed by $T_2$, for what reason we call it \lq inelastic\rq\
scattering.

\subsection{Photodetection signal}
\label{spower}

Given the final photon state $|f\rangle$,
the intensity of the photodetection signal, as measured
by a
broadband detector (polarization $\epsilon_D$) 
located at ${\bf R}$ at time $t$ reads \cite{CCT}:
\begin{equation}
I=\langle f|E^{(-)}({\bf R},t) E^{(+)}({\bf R},t) |f\rangle.
\label{intensity}
\end{equation}
Here, the detection of the photon is described by the electric field operator
\begin{equation}
E^{(+)}({\bf R},t)=\frac{g}{d}\sum_{{\bf k},s}(\epsilon_{{\bf k}s}\cdot
\epsilon_D^*)
e^{i({\bf k}\cdot{\bf R}-\omega t)}a_{{\bf k}s},\label{det}
\end{equation} 
which annihilates a photon at position ${\bf R}$.

As initial state, we consider a state of $N$ photons
\begin{equation}
|i_N\rangle=\sqrt{N!}\sum_{({\bf k}_1\dots{\bf k}_N)}h({\bf k}_1)\dots
h({\bf k}_N)|{\bf k}_1\epsilon_L,\dots,{\bf k}_N\epsilon_L\rangle,\label{i}
\end{equation} 
where all photons are described by the same single-photon wavepacket
\begin{equation}
|i_1\rangle=\sum_{\bf k}h({\bf k})|{\bf k}\epsilon_L\rangle.
\end{equation}
The factor $\sqrt{N!}$ in Eq.~(\ref{i}) arises from the symmetry under
exchange of photons as bosonic particles, and is required
to obtain the correct
normalization, $\langle i_N|i_N\rangle=\langle i_1|i_1\rangle=1$.
We assume that the wavepacket describes an
almost plane wave, i.e., $h(\bf k)$ is sharply peaked around its center
${\bf k}_L$
(\lq sharply\rq\ means: much narrower than $\Gamma$). For this reason, we
may also 
neglect in Eq.~(\ref{i}) the dependence of the initial polarization vector
$\epsilon_L$ on
$\bf k$.

The initial state $|i_N\rangle$ corresponds to the
following incident intensity seen by the atom at position 
${\bf r}$ and time $t=0$, obtained by inserting $|i_N\rangle$ instead
of $|f\rangle$ in Eq.~(\ref{intensity}), and summing over the detector
polarization $\epsilon_D$:
\begin{equation}
I_{\rm in}=
N\frac{g^2}{d^2}
\Biggl|\sum_{\bf k}e^{i{\bf k}\cdot{\bf r}}h({\bf k})\Biggr|^2.
\label{satpar2}
\end{equation}
In the following, we will use 
a dimensionless quantity, the so-called
\lq saturation parameter\rq
\begin{equation}
s=\frac{2 d^2 I_{\rm in}}{|\omega_L-\omega_0|^2}=
\frac{2Ng^2}{|\omega_L-\omega_0|^2} 
\Biggl|\sum_{\bf k}e^{i{\bf k}\cdot{\bf r}}h({\bf k})\Biggr|^2.
\label{satpar}
\end{equation} 
It accounts for the fact
that photons interact less strongly with the
atom if they are far detuned from the atomic resonance
(i.e. if $|\omega_L-\omega_0|$ is large). From the solution of the
optical Bloch equations \cite{CCT}, it is known that $s$ determines the ratio 
between inelastic and elastic scattering, see Eq.~(\ref{bloch}) below.

We are interested in the photodetection signal measured at position
${\bf R}$ at time $t=|{\bf R}-{\bf r}|$ (the time needed for the
scattered light to reach the detector, in units where $c=1$). We assume
that the detector is placed far away from the atom, such that 
$t=|{\bf R}-{\bf r}|$ is long enough for the scattering approach to be valid.
Furthermore, the detector should not be placed in the
direction of the initial wavevector ${\bf k}_L$,
such that only scattered photons are detected 
(i.e. $E^{(+)}({\bf R},t)|i_1\rangle=0$).

In order to proceed, we have to generalize
the scattering operator for two photons, Eq.~(\ref{ss2}), to the case of
$N$ photons. For this purpose, we assume that the saturation parameter $s$
is so small that
at most one photon pair is scattered
inelastically. This yields the photodetection signal up to 
second order in $s$, see below.
Summing over the different pairs $(i,j)$, and taking into
account 
all possible permutations of the $N$ photons, we obtain:
\begin{eqnarray}
\langle {\bf k}_1'\dots {\bf k}_N'|S_N|{\bf k}_1\dots {\bf k}_N\rangle & = & 
\sum_{P_N}\prod_{l=1}^N \langle {\bf k}_l'|S_1|{\bf k}_{P_N(i)}\rangle~+
\nonumber\\
& & \!\!\!\!\!\!\!\!\!\!\!\!\!\!\!\!\!\!\!\!\!\!\!\!
\!\!\!\!\!\!\!\!\!\!\!\!\!\!\!\!\!\!\!\!\!\!\!\!
\!\!\!\!\!\!\!\!\!\!\!\!\!\!\!\!\!\!\!
+\sum_{\stackrel{\scriptstyle i,j=1}{i<j}}^N\sum_{P_N/P_2}
\langle {\bf k}_i'{\bf k}_j'|T_2|
{\bf k}_{P_N(i)}{\bf k}_{P_N(j)}\rangle\prod_{\stackrel{\scriptstyle l=1}
{l\neq i,j}}^N\langle{\bf k}_l'|S_1|{\bf k}_{P_N(l)}\rangle.
\label{sn}
\end{eqnarray}
(In the following, we do not write explicitly the polarization vectors.)
Eq.~(\ref{sn}) contains 
a sum over all permutations $P_N$ of the $N$ indices 
$\{1\dots N\}$, modulo a permutation of the two indices $P_N(i)$ and
$P_N(j)$ in the second term, where the latter permutation is included in
the two-photon operator $T_2$, see Eq.~(\ref{t2}). 
In the case $N=2$, the above expression agrees with the one of the
previous section, Eq.~(\ref{ss2}). According to Eq.~(\ref{sn}),
the final photon state $|f_N\rangle=S_N|i_N\rangle$ can be expressed as
follows:
\begin{eqnarray}
\langle {\bf k}_1\dots{\bf k}_N|f_N\rangle & = & 
\sqrt{N!}\prod_{l=1}^N \langle {\bf k}_l|f_1\rangle+
\nonumber\\ & + & \sqrt{\frac{N!}{2}}
\sum_{\stackrel{\scriptstyle i,j=1}{i<j}}^N\langle {\bf k}_i{\bf k}_j|g_2
\rangle
\prod_{\stackrel{\scriptstyle l=1}
{l\neq i,j}}^N\langle {\bf k}_l|f_1\rangle,\label{fn}
\end{eqnarray}
in terms of the one- and two-photon states
\begin{eqnarray}
|f_1\rangle & = & S_1|i_1\rangle,\label{f1}\\
|g_2\rangle & = & T_2|i_2\rangle.\label{g2}
\end{eqnarray}

Following
Eq.~(\ref{intensity}), we now apply the electric field operator on the
final photon state. It may annihilate either an elastically or an
inelastically scattered photon. Correspondingly, we obtain the
following three contributions:
\begin{equation}
|\psi\rangle=E^{(+)}({\bf R},t)|f\rangle=
\sum_{i=1}^3 |\psi_i\rangle,\label{psi}
\end{equation}
with
\begin{eqnarray}
\langle {\bf k}_1\dots{\bf k}_{N-1}|\psi_1\rangle & = & 
\sqrt{N!} E \prod_{l=1}^{N-1}\langle {\bf k}_l|f_1\rangle,\label{psi1}\\
\langle {\bf k}_1\dots{\bf k}_{N-1}|\psi_2\rangle & = & 
\sqrt{\frac{N!}{2}}\sum_{i=1}^{N-1}\langle {\bf k}_i|g_1\rangle
\prod_{\stackrel{\scriptstyle l=1}
{l\neq i}}^{N-1}\langle {\bf k}_l|f_1\rangle,\label{psi2}\\
\langle {\bf k}_1\dots{\bf k}_{N-1}|\psi_3\rangle & = & 
\sqrt{\frac{N!}{2}} E \sum_{\stackrel{\scriptstyle i,j=1}{i<j}}^{N-1}
\langle {\bf k}_i{\bf k}_j|g_2\rangle
\prod_{\stackrel{\scriptstyle l=1}
{l\neq i,j}}^{N-1}\langle {\bf k}_l|f_1\rangle,\label{psi3}
\end{eqnarray}
and
\begin{eqnarray}
E & = & \langle 0|E^{(+)}({\bf R},t)|f_1\rangle,\label{e}\\
|g_1\rangle & = & E^{(+)}({\bf R},t)|g_2\rangle.\label{g1}
\end{eqnarray}
According to Eq.~(\ref{intensity}), the norm 
$I=\langle\psi|\psi\rangle$ gives the
total intensity. Let us first concentrate on the contributions from
$|\psi_1\rangle$ and $|\psi_2\rangle$. (As we will argue later,
$|\psi_3\rangle$ can be neglected).
We obtain a sum of three terms, from elastic and
inelastic scattering, and their interference. Using
$\langle f_1|f_1\rangle=1$ (since $S_1$ is unitary), we obtain:
\begin{eqnarray}
I_{\rm el}^{(1)} & = & \langle\psi_1|\psi_1\rangle=N|E|^2,\label{i1}\\
I_{\rm el}^{(2)} & = & \langle\psi_1|\psi_2\rangle+\langle\psi_2|\psi_1\rangle
\nonumber\\
& = & 
N(N-1){\rm Re}\{\sqrt{2}E\langle g_1|f_1\rangle\},
\label{i2}\\
I_{\rm in} & = & \langle\psi_2|\psi_2\rangle=
\frac{N(N-1)}{2}
\langle g_1|g_1\rangle\nonumber\\
& & + \frac{N(N-1)(N-2)}{2}|\langle f_1|g_1\rangle|^2.\label{i3}
\end{eqnarray}
Whereas in $I_{\rm el}^{(1,2)}$, the frequency of the detected photon
is fixed to $\omega=\omega_L$ (since one-photon scattering is elastic),
this is not the case for $I_{\rm in}$, where the overlap
$\langle g_1|g_1\rangle$ implies an integral over $\omega$. Thereby,
we obtain an elastic and inelastic component of the detection signal.
 
To complete the calculation, we insert the one- and
two-photon scattering matrices given in Sec.~\ref{scatter1}.
Using Eqs.~(\ref{ss1},\ref{t1}), the final one-photon state reads:
\begin{eqnarray}
|f_1\rangle & = & |i_1\rangle~-~\frac{2\pi ig^2}{\omega_L-\omega_0}\times
\nonumber\\& & \!\!\!\!\!\!\!\!\!\!\!\!
\times\sum_{{\bf k}_i,{\bf k}_f\epsilon_f} h({\bf k}_i)
\delta(\omega_i-\omega_f)(\epsilon_L\cdot\epsilon_f^*)
 e^{i({\bf k}_i-{\bf k}_f)\cdot{\bf r}} 
|{\bf k}_f\epsilon_f\rangle.
\end{eqnarray}
Since the wavepacket $h({\bf k}_i)$ is quasi-monochromatic, we may
replace the argument of functions which vary
slowly (i.e. on the scale of $\Gamma$) as a function of $\omega_i$
by the constant value $\omega_L$. Applying the electric field operator
on $|f_1\rangle$,
see Eq.~(\ref{e}),
yields, under the assumptions given above:
\begin{equation}
E = \frac{-3\Gamma(\epsilon_L\cdot\epsilon_D^*) g}{4\omega_L d R
(\omega_L-\omega_0)}
\sum_{{\bf k}_i}e^{i{{\bf k}_i}\cdot{\bf r}}
h({\bf k}_i),
\end{equation}
Similarly, we obtain
for the inelastic part, see Eqs.~(\ref{t2},\ref{g2},\ref{g1}):
\begin{equation}
|g_1\rangle = - 2 E g^2 \sum_{{\bf k}_i,{\bf k}_f \epsilon_f}
\frac{\sqrt{2}e^{i({\bf k}_i-{\bf k}_f)\cdot{\bf r}}
h({\bf k}_i)(\epsilon_L\cdot\epsilon_f^*)}
{(\omega_f-\omega_0)(2\omega_L-\omega_f-\omega_0)}|{\bf k}_f\epsilon_f\rangle.
\end{equation}
According to Eqs.~(\ref{i1}-\ref{i3}), we obtain the following
intensity $I=I_{\rm el}^{(1)}+I_{\rm el}^{(2)}+I_{\rm in}$ of the
photodetection signal:
\begin{eqnarray}
I_{\rm el}^{(1)} & = & \eta \frac{s}{2},\ \ 
I_{\rm el}^{(2)} = -\eta \frac{N-1}{N}s^2,\label{el}\\ 
I_{\rm in} & = & \eta \frac{N-1}{N}\frac{s^2}{2}~+~o(s^3)\label{in},
\end{eqnarray}
with the prefactor:
\begin{equation}
\eta=\left(\frac{3\Gamma|\epsilon_L\cdot\epsilon_D^*|}{4d\omega_L R}\right)^2.
\label{eta}
\end{equation}
The term proportional to $|\langle f_1|g_1\rangle|^2$
in Eq.~(\ref{i3}) gives a contribution to the
inelastic component in third order of $s$, which can be neglected. 
As it should be, for large $N$ - such that the first photon can be absorbed
without significantly changing the saturation induced by the remaining
$(N-1)$ photons - the above result agrees with 
the elastic and inelastic components
\begin{equation}
I_{\rm el}=\eta \frac{s}{2(1+s)^2},\ \ I_{\rm in}=\eta \frac{s^2}{2(1+s)^2}
\label{bloch}
\end{equation}
of the resonance fluorescence as predicted by the Bloch equations \cite{CCT},
expanded up to second order in $s$. 

However, we have not yet accounted for the third term $|\psi_3\rangle$ in
Eq.~(\ref{psi}). If we compare Eqs.~(\ref{fn},\ref{psi1},\ref{psi3}),
we note that $|\psi_1\rangle+|\psi_3\rangle=\sqrt{N}E|f_{N-1}\rangle$,
and hence the norm of $|\psi_1\rangle+|\psi_3\rangle$ equals the
norm of $|\psi_1\rangle$, Eq.~(\ref{i1}), provided
that the norm of $|f_{N-1}\rangle$ is 1. Although the latter condition is
not necessarily fulfilled if the scattering operator is truncated 
as in Eq.~(\ref{sn}), its unitarity will be recovered when including
higher scattering orders. Similarly, it can be shown that contributions
from $\langle\psi_2|\psi_3\rangle$ - if they are not of third order in $s$ -
are exactly canceled by other terms
which appear in $\langle\psi_2|\psi_1\rangle$
when including into $\langle\psi_2|$ another inelastically scattered
photon pair. Hence, the term $|\psi_3\rangle$ does not contribute to the
photodetection signal up to second order in the saturation parameter $s$.   

By putting a spectral filter in front of the detector, we can resolve
the power spectrum $P(\omega)$ of the detection signal, i.e. the probability 
of detecting a photon of a definite frequency $\omega$. Since elastic
scattering conserves the frequency, the spectrum exhibits a sharp peak
at $\omega_L$ (almost a $\delta$-function for our quasi-monochromatic
initial wavepacket $f$),
\begin{equation}
P(\omega)=I^{(\rm el)}\delta_f(\omega-\omega_L)+P^{(\rm in)}(\omega),
\label{spectotal}
\end{equation}
whereas the inelastic component depends smoothly
on $\omega$. The latter is
proportional to the absolute square of the 
\begin{figure}
\centerline{\epsfig{figure=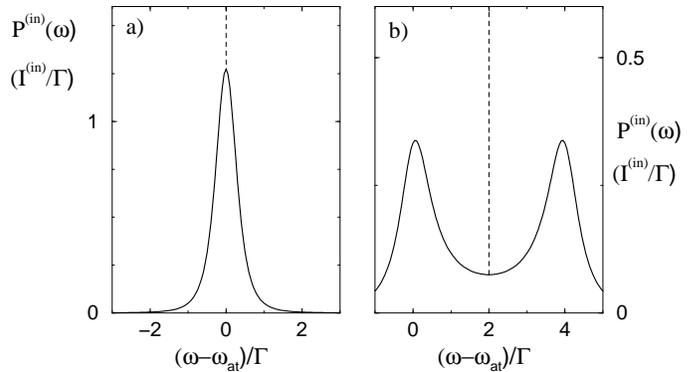,height=9cm,angle=270}}
\caption{Inelastic resonance fluorescence
spectrum $P^{(\rm in)}(\omega)$, Eq.~(\ref{specin}),
for small saturation, $s\ll 1$, (a) zero detuning 
$\delta=\omega_L-\omega_{\rm at}=0$,
and (b) $\delta=2\Gamma$. The dashed lines indicate the position of the
elastic peak at $\omega_L$, see Eq.~(\ref{spectotal}).}
\label{spec}
\end{figure}
inelastic transition amplitude,
Eq.~(\ref{t2}) (with $\omega_1=\omega_2=\omega_L$ the initial frequency,
$\omega_3=\omega$ the frequency of the detected photon, and
$\omega_4=2\omega_L-\omega$). With the correct normalization, 
\begin{equation}
I^{\rm (in)}=\int d\omega P^{\rm (in)}(\omega),\label{inintegral}
\end{equation} 
we obtain:
\begin{equation}
P^{(\rm in)}(\omega) = \frac{\Gamma I^{(\rm in)}}{4\pi}
\left|\frac{1}{\omega-\omega_0}+
\frac{1}{2\omega_L-\omega-\omega_0}\right|^2.\label{specin}
\end{equation}
For zero detuning, 
$\omega_L=\omega_{\rm at}$, the inelastic spectrum consists of a peak of width
$0.64\Gamma$, whereas for large detuning 
$\delta=\omega_L-\omega_{\rm at}$ (i.e. if $4\delta^2\gg\Gamma^2$),
there are two peaks of width $\Gamma$ 
at $\omega=\omega_L\pm\delta$, see Fig.~\ref{spec}. 
\footnote{The reader may have in mind that the resonance fluorescence
actually exhibits {\em three} peaks \cite{mollow}. However, the
one at $\omega=\omega_L$ is of higher order in $s$, since it arises from
three-photon scattering.}
Note that one of them
is centered exactly at the atomic resonance. Evidently,
this will be important
if we allow the scattered photons to interact with a second atom,
as we will do now.

\section{Two atoms}
\label{double}

\subsection{Scattering matrix}
\label{scatter2}

Let us now turn to the case of two atoms alone in vacuum.
We assume that the second atom is far away from the first one,
compared to the optical wavelength. This means that we may restrict
ourselves to processes where at most one of the two
photons is scattered by both atoms. As shown in appendix~\ref{app2at},
the corresponding scattering matrix can then 
be obtained in a simple way
from the single-atom scattering matrix, see Eqs.~(\ref{a1},\ref{a2}):
apart from the
geometrical phase factors $e^{\pm i{\bf k}\cdot {\bf r}_{1,2}}$
for absorption or emission of a photon $|{\bf k}\rangle$ by atom 1 or 2,
and the terms depending on the polarization,
we only have to take into account the \lq photon exchange factor\rq
\begin{equation}
B(\omega)=-\frac{3\Gamma e^{i\omega r_{12}}}{4\omega r_{12}
(\omega-\omega_0)},\label{beta}
\end{equation}
depending on the frequency $\omega$ of the doubly scattered photon.
In Eq.~(\ref{beta}),
we recognize the propagation of a spherical wave from
one atom to the other one
(inversely proportional to their distance $r_{12}$), and the
amplitude $(\omega-\omega_0)^{-1}$ describing scattering by a single atom,
see Eq.~(\ref{t1}). 

In particular, the structure of the scattering operator as a sum of
an elastic single-photon and an inelastic two-photon component is
the same as before, compare Eqs.~(\ref{ss1},\ref{ss2}):
\begin{eqnarray}
\langle {\bf k_3}\epsilon_3,{\bf k_4}\epsilon_4|S_2^{(2)}|{\bf k_1}\epsilon_1,{\bf k_2}\epsilon_2\rangle & = &\nonumber\\
& & \!\!\!\!\!\!\!\!\!\!\!\!\!\!\!\!\!\!\!\!\!\!\!\!\!\!\!\!\!\!\!\!
\!\!\!\!\!\!\!\!\!\!\!\!\!\!\!\!\!\!\!\!\!\!\!\!\!\!\!\!\!\!\!\!
\langle {\bf k_3}\epsilon_3|S_1^{(2)}|{\bf k_1}\epsilon_1\rangle\langle {\bf k_4}\epsilon_4|S_1^{(2)}|{\bf k_2}\epsilon_2\rangle
\nonumber\\
& & \!\!\!\!\!\!\!\!\!\!\!\!\!\!\!\!\!\!\!\!\!\!\!\!\!\!\!\!\!\!\!\! +
\langle {\bf k_3}\epsilon_3|S_1^{(2)}|{\bf k_2}\epsilon_2\rangle\langle
{\bf k_4}\epsilon_4|S_1^{(2)}|{\bf k_1}\epsilon_1\rangle\nonumber\\
& & \!\!\!\!\!\!\!\!\!\!\!\!\!\!\!\!\!\!\!\!\!\!\!\!\!\!\!\!\!\!\!\!
\!\!\!\!\!\!\!\!\!\!\!\!\!\!\!\!\!\!\!\!\!\!\!\!\!\!\!\!\!\!\!\!
+ \langle {\bf k_3}\epsilon_3,{\bf k_4}\epsilon_4|T_2^{(2)}|
{\bf k_1}\epsilon_1,{\bf k_2}\epsilon_2\rangle,
\label{prod2}
\end{eqnarray}
where the single-photon component of $S^{(2)}$ 
contains also the non-scattered wave,
see diagram (1a) in Fig.~\ref{2atom1}:
\begin{equation}
S_1^{(2)}={\mathbbm 1}-2\pi i\delta(\omega_f-\omega_i) T_1^{(2)}.
\end{equation}
The remaining single-photon processes are also shown in
Fig.~\ref{2atom1}. The photon may be scattered by only one
atom (1 or 2), or by both (first 1, then 2, and vice versa). Correspondingly,
the single-photon transition operator reads:
\begin{eqnarray}
\langle {\bf k}_f\epsilon_f|T_1^{(2)}|{\bf k}_i\epsilon_i\rangle & = &
\frac{g^2}{\omega_i-\omega_0}\times\Biggl\{
\Biggr.
\nonumber\\ 
& & \!\!\!\!\!\!\!\!\!\!\!\!\!\!\!\!\!\!\!\!\!\!\!\!\!\!\!\!\!\!\!\!\!\!\!\!\!
\!\!\!\!\!\!\!\!\!\!\!
e^{i({\bf k}_i-{\bf k}_f)\cdot{\bf r}_1}\Bigl(
(\epsilon_i\cdot\epsilon_f^*)+
B(\omega_i)(\epsilon_i\cdot\Delta_{12}\cdot\epsilon_f^*)
e^{i{\bf k}_i\cdot({\bf r}_2-{\bf r}_1)}\Bigr)\nonumber\\
& &  \!\!\!\!\!\!\!\!\!\!\!\!\!\!\!\!\!\!\!\!\!\!\!\!\!\!\!\!\!\!\!\!\!\!\!\!\!
\!\!\!\!\!\!\!\!\!\!\!
\Biggl. e^{i({\bf k}_i-{\bf k}_f)\cdot{\bf r}_2}\Bigl(
(\epsilon_i\cdot\epsilon_f^*)+
B(\omega_i)(\epsilon_i\cdot\Delta_{12}\cdot\epsilon_f^*)
e^{i{\bf k}_i\cdot({\bf r}_2-{\bf r}_1)}\Bigr)
\Biggr\}
\label{t12}
\end{eqnarray}
As mentioned above, for the two double-scattering
processes, see diagram (1d) and (1e) in Fig.~\ref{2atom1}, we have to
multiply the one-atom transition operator
$\langle {\bf k}_f\epsilon_f|T|{\bf k}_i\epsilon_i\rangle$, Eq.~(\ref{t1}),
with the photon exchange factor $B(\omega_i)$,
see Eq.~(\ref{beta}), and to adjust the geometrical phase factor.
Furthermore, 
the fact that the photon propagates 
in the direction ${\bf r}_2-{\bf r}_1$ between 
the two scattering events implies
a projection $\Delta_{12}$ of the polarization vector
onto the plane perpendicular to ${\bf r}_2-{\bf r}_1$.
Thereby, the term $\epsilon_i\cdot\epsilon_f^*$
(for scattering by a single atom) is replaced by
$\epsilon_i\cdot\Delta_{12}\cdot\epsilon_f^*$.

In the case of inelastic two-photon scattering, the doubly scattered
photon may be scattered first inelastically (by atom 1 or 2), and then
elastically (by the other atom), or vice versa, compare, e.g., the
diagrams (2a) and (2d) in Fig.~\ref{2atom2}.
Correspondingly, the frequency to be inserted in the photon exchange factor
$B(\omega)$, Eq.~(\ref{beta}), is either the final or initial
frequency of this photon, see Eq.~(\ref{a1}) or Eq.~(\ref{a2}).
In total, we obtain:
\begin{eqnarray}
\langle {\bf k}_3\epsilon_3,{\bf k}_4\epsilon_4|T_2^{(2)}|
{\bf k}_1\epsilon_1,{\bf k}_2\epsilon_2\rangle & = &\nonumber\\
& & \!\!\!\!\!\!\!\!\!\!\!\!\!\!\!\!\!\!\!\!\!\!\!\!\!\!\!\!\!\!\!\!\!\!\!\!\!
\!\!\!\!\!\!\!\!\!\!\!\!\!\!\!\!\!\!\!\!\!\!\!\!\!\!\!\!\!\!\!\!\!\!\!\!\!\!\!
2\pi i
\frac{g^4 \delta(\omega_1+\omega_2-\omega_3-\omega_4)} 
{(\omega_1-\omega_0)(\omega_2-\omega_0)}
\left(\frac{1}{\omega_3-\omega_0}+\frac{1}{\omega_4-\omega_0}\right)\nonumber\\
& & \!\!\!\!\!\!\!\!\!\!\!\!\!\!\!\!\!\!\!\!\!\!\!\!\!\!\!\!\!\!\!\!\!\!\!\!\!
\!\!\!\!\!\!\!\!\!\!\!\!\!\!\!\!\!\!\!\!\!\!\!\!\!\!\!\!\!\!\!\!\!\!
e^{i({\bf k}_1+{\bf k}_2-{\bf k}_3-{\bf k}_4)\cdot{\bf r}_1}\times
\Biggr\{(\epsilon_1\cdot\epsilon_3^*)(\epsilon_2\cdot\epsilon_4^*)
\Biggl.\nonumber\\
& & \!\!\!\!\!\!\!\!\!\!\!\!\!\!\!\!\!\!\!\!\!\!\!\!\!\!\!\!\!\!\!\!\!\!\!\!\!
\!\!\!\!\!\!\!\!\!\!\!\!\!\!\!\!\!+~
B(\omega_1)~(\epsilon_1\cdot\Delta_{12}\cdot\epsilon_3^*)
(\epsilon_2\cdot\epsilon_4^*)~
e^{i{\bf k}_1\cdot({\bf r}_2-{\bf r}_1)}
\nonumber\\
& & \!\!\!\!\!\!\!\!\!\!\!\!\!\!\!\!\!\!\!\!\!\!\!\!\!\!\!\!\!\!\!\!\!\!\!\!\!
\!\!\!\!\!\!\!\!\!\!\!\!\!\!\!\!\!+~
B(\omega_2)~(\epsilon_1\cdot\epsilon_3^*)(\epsilon_2\cdot\Delta_{12}\cdot
\epsilon_4^*)~
e^{i{\bf k}_2\cdot({\bf r}_2-{\bf r}_1)}
\nonumber\\
& & \!\!\!\!\!\!\!\!\!\!\!\!\!\!\!\!\!\!\!\!\!\!\!\!\!\!\!\!\!\!\!\!\!\!\!\!\!
\!\!\!\!\!\!\!\!\!\!\!\!\!\!\!\!\!+
~B(\omega_3)~(\epsilon_1\cdot\Delta_{12}\cdot\epsilon_3^*)
(\epsilon_2\cdot\epsilon_4^*)~
e^{-i{\bf k}_3\cdot({\bf r}_2-{\bf r}_1)}
\nonumber\\
& & \!\!\!\!\!\!\!\!\!\!\!\!\!\!\!\!\!\!\!\!\!\!\!\!\!\!\!\!\!\!\!\!\!\!\!\!\!
\!\!\!\!\!\!\!\!\!\!\!\!\!\!\!\!\!\Biggl.+
~B(\omega_4)~(\epsilon_1\cdot\epsilon_3^*)(\epsilon_2\cdot\Delta_{12}\cdot
\epsilon_4^*)~
e^{-i{\bf k}_4\cdot({\bf r}_2-{\bf r}_1)}
\Biggr\}\nonumber\\
& & \!\!\!\!\!\!\!\!\!\!\!\!\!\!\!\!\!\!\!\!\!\!\!\!\!\!\!\!\!\!\!\!\!\!\!\!\!
\!\!\!\!\!\!\!\!\!\!\!\!\!\!\!\!\!\!\!\!\!\!\!\!\!\!\!\!\!\!\!\!\!\!
+~e^{i({\bf k}_1+{\bf k}_2-{\bf k}_3-{\bf k}_4)\cdot{\bf r}_2}\times
\Biggr\{(\epsilon_1\cdot\epsilon_3^*)(\epsilon_2\cdot\epsilon_4^*)
\Biggl.\nonumber\\
& & \!\!\!\!\!\!\!\!\!\!\!\!\!\!\!\!\!\!\!\!\!\!\!\!\!\!\!\!\!\!\!\!\!\!\!\!\!
\!\!\!\!\!\!\!\!\!\!\!\!\!\!\!\!\!+~
B(\omega_1)~(\epsilon_1\cdot\Delta_{12}\cdot\epsilon_3^*)
(\epsilon_2\cdot\epsilon_4^*)~
e^{i{\bf k}_1\cdot({\bf r}_1-{\bf r}_2)}
\nonumber\\
& & \!\!\!\!\!\!\!\!\!\!\!\!\!\!\!\!\!\!\!\!\!\!\!\!\!\!\!\!\!\!\!\!\!\!\!\!\!
\!\!\!\!\!\!\!\!\!\!\!\!\!\!\!\!\!+~
B(\omega_2)~(\epsilon_1\cdot\epsilon_3^*)(\epsilon_2\cdot\Delta_{12}\cdot
\epsilon_4^*)~
e^{i{\bf k}_2\cdot({\bf r}_1-{\bf r}_2)}
\nonumber\\
& & \!\!\!\!\!\!\!\!\!\!\!\!\!\!\!\!\!\!\!\!\!\!\!\!\!\!\!\!\!\!\!\!\!\!\!\!\!
\!\!\!\!\!\!\!\!\!\!\!\!\!\!\!\!\!+
~B(\omega_3)~(\epsilon_1\cdot\Delta_{12}\cdot\epsilon_3^*)
(\epsilon_2\cdot\epsilon_4^*)~
e^{-i{\bf k}_3\cdot({\bf r}_1-{\bf r}_2)}
\nonumber\\
& & \!\!\!\!\!\!\!\!\!\!\!\!\!\!\!\!\!\!\!\!\!\!\!\!\!\!\!\!\!\!\!\!\!\!\!\!\!
\!\!\!\!\!\!\!\!\!\!\!\!\!\!\!\!\!\Biggl.+
~B(\omega_4)~(\epsilon_1\cdot\epsilon_3^*)(\epsilon_2\cdot\Delta_{12}\cdot
\epsilon_4^*)~
e^{-i{\bf k}_4\cdot({\bf r}_1-{\bf r}_2)}
\Biggr\}\nonumber\\
& & \!\!\!\!\!\!\!\!\!\!\!\!\!\!\!\!\!\!\!\!\!\!\!\!\!\!\!\!\!\!\!\!\!\!\!\!\!
\!\!\!\!\!\!\!\!\!\!\!\!\!\!\!\!\!\!\!\!\!\!\!\!\!\!\!\!\!\!\!\!\!\!\!\!\!\!\!
+~\Bigl({\bf k}_1\epsilon_1\leftrightarrow
{\bf k}_2\epsilon_2\Bigr)
\label{t22}
\end{eqnarray}
Here, the last line denotes additional terms arising from exchanging
the initial (or, equivalently, final) photons.
We recognize two terms describing the scattering
by atom 1 or 2 alone, see diagrams (2i) and (2j) in Fig.~\ref{2atom3},
and eight different terms describing the processes where both atoms
are involved, see diagrams (2a-2h) in Figs.~\ref{2atom2} and
\ref{2atom3}.
Note that the terms depending on the polarization allow to
identify the photon which is scattered by both atoms.
This photon is marked by full arrows in
Figs.~\ref{2atom1}-\ref{2atom3}, whereas the open arrows denote the
photon scattered by only one atom. If we assume that
$|{\bf k}_4\epsilon_4\rangle$ is the doubly scattered photon,
the ten terms in Eq.~(\ref{t22}) correspond (from top to bottom)
to the diagrams
(2i), (2e), (2a), (2h), (2d), 
(2j), (2f), (2b), (2g), and (2c), respectively.

\subsection{Direct calculation of the enhancement factor}
\label{backscattering}

Having at hand the scattering matrix, we now determine the intensity
of the photodetection signal. In principle,
the calculation can be performed in the same
way as in the single-atom case, Sec.~\ref{spower}. However, the detection
signal will now depend non-trivially on the position of the atoms and
the detector, due to the fact that 
\begin{figure}[t]
\centerline{\epsfig{figure=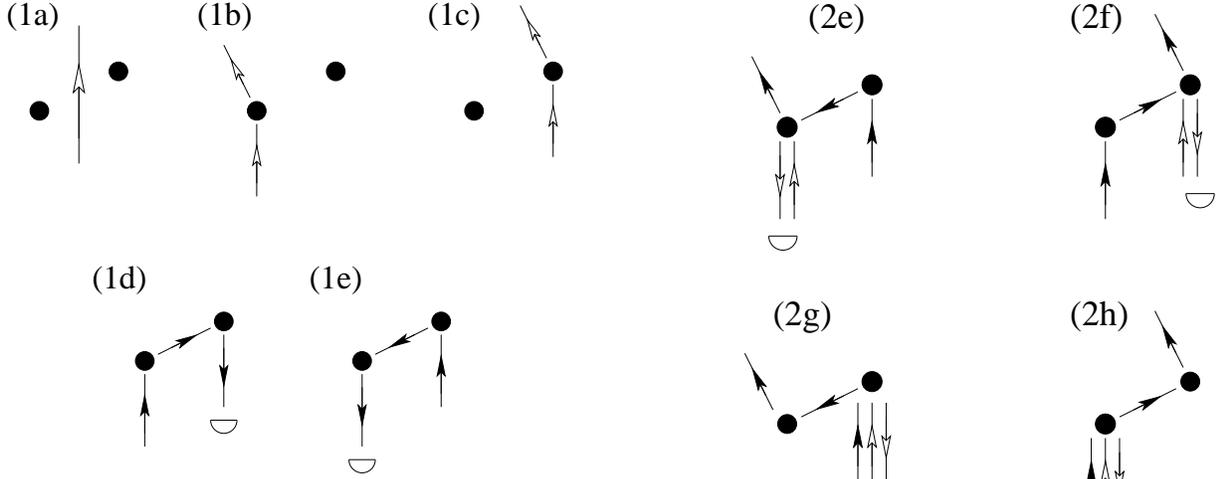,width=8cm,angle=0}}
\caption{Scattering of a single photon by two distant
atoms. In the coherent backscattering experiment,
only the doubly scattered photon is detected, see diagrams (1d,e).
Consequently, the diagrams (1a-c), with single or no scattering,
describe the undetected photon.}
\label{2atom1}
\end{figure}
\begin{figure}[t]
\centerline{\epsfig{figure=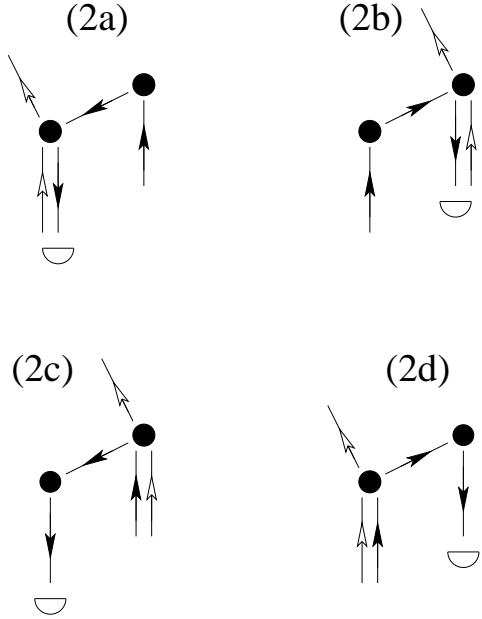,width=7cm,angle=0}}
\caption{Inelastic scattering of two photons by two distant
atoms. Only the doubly scattered photon (full arrows) is detected.
Since the photon frequencies are changed by the inelastic scattering
event at the atom where both photons meet, the amplitude
of the elastic scattering event at the second atom
depends on whether the
inelastic scattering occurs before the elastic one (2c,d) or after
(2a,b).} 
\label{2atom2}
\end{figure}
the photons emitted by one atom 
interfere with the photons emitted by the other one. Even if we average over
the positions ${\bf r}_1$ and ${\bf r}_2$ of the atoms, the interference is
not completely washed out. There remains an 
\begin{figure}[t]
\centerline{\epsfig{figure=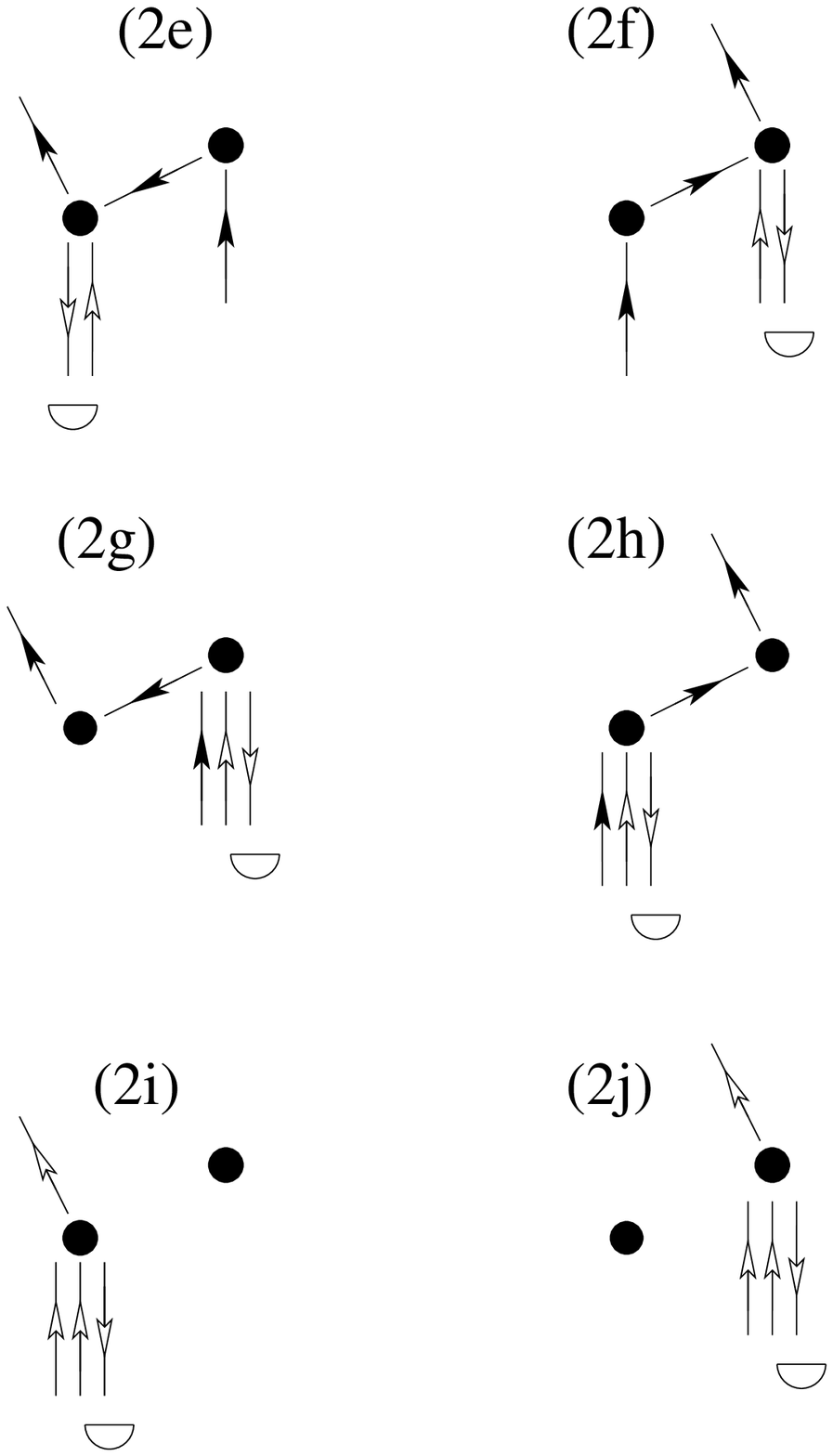,width=7cm,angle=0}}
\caption{Remaining diagrams describing inelastic scattering of two photons
by two atoms. In the coherent backscattering experiment, they are filtered
out by using the $h\parallel h$ polarization channel
(see Sec.~\ref{backscattering}),
in which a singly scattered photon (open arrows) cannot be detected.}
\label{2atom3}
\end{figure}
enhanced probability
to detect a photon in the direction opposite to the incident wave - an effect
which is known as {\em coherent backscattering}. In the case of two
atoms, it arises from double scattering: in the backscattering
direction, a photon scattered first by atom 1, and then by atom 2,
interfers constructively with the corresponding reversed path.

In order to examine cleanly this interference effect,
we therefore assume that only doubly scattered photons are detected.
Experimentally, this can be realized by using circularly polarized
light $\epsilon_L=(1,i,0)/\sqrt{2}$ (in Euclidean coordinates, where the
$z$ axis is parallel to ${\bf k}_L$)
, and detecting the scattered photons in the
helicity preserving channel $\epsilon_D=\epsilon_L^*$.
This implies $(\epsilon_L\cdot\epsilon_D^*)=0$, i.e.
no singly scattered photons
can be detected in the helicity preserving polarization channel.
If we look at the inelastic part of the scattering matrix, Eq.~(\ref{t22}),
assuming (without loss of generality) that the
photon $|{\bf k}_4\epsilon_4\rangle$ is the detected one,
this means that all terms with $(\epsilon_{1}\cdot\epsilon_4^*)$ or
$(\epsilon_{2}\cdot\epsilon_4^*)$ are filtered out. These are the diagrams 
shown in Fig.~\ref{2atom3}, and only those of Fig.~\ref{2atom2}
remain.  
Concerning the elastic single-photon 
scattering, see Fig.~\ref{2atom1} we keep the single-scattering 
diagrams (1a-c) to describe the undetected photon. 
For the sake of completeness, we will repeat in appendix~\ref{scalar} 
the following calculation for the case of scalar photons,
where, a priori, all the diagrams shown in Figs.~\ref{2atom1}-\ref{2atom3}
contribute.

\subsubsection{Elastic contribution}

Let us begin with the contribution $I_{\rm el}^{(1)}$ of one-photon scattering.
According to Eqs.~(\ref{e},\ref{i1}), it is obtained by applying the
electric field on the final state $|f_1\rangle$ of single-photon
scattering. As explained above, only
the diagrams (1d) and (1e) in Fig.~\ref{2atom1} contribute.
At first, we concentrate on the 
phase factors depending on the position of the atoms.
If ${\bf k}_L$ is the wavevector of
the incident photon, and the detector is located in the
direction ${\bf k}_D$ (with $|{\bf k}_D|=|{\bf k}_L|$,
since one-photon scattering conserves the frequency), we obtain
$\exp(i{\bf r}_1\cdot{\bf k}_L-i{\bf r}_2\cdot{\bf k}_D)$ for (1d) 
and $\exp(i{\bf r}_2\cdot{\bf k}_L-i{\bf r}_1\cdot{\bf k}_D)$
for (1e). Evidently, the phases are identical if ${\bf k}_D=-{\bf k}_L$, i.e.,
(1d) and (1e) interfere constructively in the backscattering
direction. On the other hand, if ${\bf k}_L\neq {\bf k}_D$ (more 
precisely: if the angle between  ${\bf k}_L$ and ${\bf k}_D$ is much larger
than some characteristic quantity $\theta_C$), the interference between
(1d) and (1e) vanishes when averaging over the positions of
the atoms. 
For simplicity, we fix the distance $r_{12}$
and average only over the angular variables of ${\bf r}_1-{\bf r}_2$.
In this case, the width of the enhanced backscattering signal
(which is also called \lq the cone\rq)
is given by $\theta_C=1/(\omega r_{12})$. In total, we obtain both for
the background intensity (known in the literature as the \lq ladder term\rq\
$L$), and
the additional intensity in backscattering direction (the \lq
crossed term\rq\ $C$) twice the result  $\eta s/2$ of the single-atom case,
\begin{equation}
L^{(1)} = C^{(1)} ={\tilde{\eta}} s,
\label{elladder1}\label{elcrossed1}
\end{equation}  
apart from a modification of the prefactor
\begin{eqnarray}
\tilde{\eta} & = & \left(\frac{3\Gamma}{4d\omega_L R}\right)^2 \langle
|B(\omega_L)|^2 |\epsilon_L\cdot
\Delta_{12}\cdot\epsilon_D^*|^2\rangle_{{\bf r}_{1,2}}\label{eta2a}\\
& = & \frac{3}{8}\left(\frac{9\Gamma^2}{16 d\omega_L^2Rr_{12}
|\omega_L-\omega_0|}\right)^2.
\label{eta2}
\end{eqnarray}
Here, Eq.~(\ref{eta2a}) implies an average over the positions of the two atoms.
The polarization-dependent term
$|\epsilon_L\cdot\Delta_{12}\cdot\epsilon_D^*|^2=\sin^4\theta/4$ is given
by the
angle $\theta$ between the incident laser ${\bf k}_L$ and the
two atoms ${\bf r}_{12}={\bf r}_1-{\bf r}_2$. Then, 
a spherical distribution of ${\bf r}_{12}$, at fixed distance $r_{12}$,
yields the result given in Eq.~(\ref{eta2}).
The fact that $L^{(1)}=C^{(1)}$ can be traced
back to the reciprocity symmetry \cite{vanT97}.

Next, we examine the interference between two-photon and one-photon
scattering, which gives rise to the elastic component $I_{\rm el}^{(2)}$ 
of the intensity
in second order of $s$, see Sec.~\ref{spower}. According to
Eq.~(\ref{i2}), $I_{\rm el}^{(2)}$ is given by the overlap of the
respective quantum states $|f_1\rangle$ and  $|g_1\rangle$ of the
undetected
photon, which amounts to a sum over the latter's state
$|{\bf k}\epsilon\rangle$ (i.e., $\langle g_1|f_1\rangle=
\sum_{{\bf k},\epsilon}\langle g_1|
{\bf k}\epsilon\rangle\langle {\bf k}\epsilon|f_1\rangle$).
First, we concentrate on 
the phase factor $\exp(-i{\bf k}\cdot {\bf r}_{1,2})$
of the undetected photon, depending
on whether it is emitted by atom 1 or 2. Integrating over the angular
variables $\Omega_{\bf k}$ of $\bf k$ (at fixed $|{\bf k}|=\omega_L$), we 
obtain, if 
$|{\bf k}\rangle$ is emitted by different atoms:
\begin{equation}
\int d\Omega_{\bf k} e^{\pm i{\bf k}\cdot ({\bf r}_1-{\bf r}_2)}=4\pi
\frac{\sin(\omega_L r_{12})}{\omega_L r_{12}}\ll 1.\label{angint}
\end{equation}    
Since we have assumed $\omega_L r_{12}\gg 1$, the above term can be neglected.
In other words, diagrams
where the undetected photon is emitted by different atoms do not interfere
in leading order of $1/(\omega_L r_{12})$.
If we now select one of the four diagrams
(2a-d) describing two-photon scattering, we can
discard among the three one-photon diagrams (1a-c),
the one where the
undetected photon
is scattered by the \lq wrong\rq\ atom. The remaining two exactly
give the final state of a photon scattered by a single atom,
as described by Eq.~(\ref{f1}). Concerning the detected photon of the
one-photon scattering, we can choose either diagram (1d) or (1e).
As already
discussed above, one of them gives a contribution to the background $L$,
and the other one to the enhanced
backscattering signal $C$. As there are in total
four diagrams (2a-d),
we obtain both for $L$ and $C$ four times the
result $-\eta s^2$ of the single-atom case:
\begin{equation}
L^{(2,\rm el)} = C^{(2,\rm el)} =
-4{\tilde{\eta}} s^2,\label{elladder2}\label{elcrossed2}
\end{equation}
Note that the total elastic ladder term,
Eq.~(\ref{elladder1}) and Eq.~(\ref{elladder2}), equals the total
elastic crossed one, Eq.~(\ref{elcrossed1}) and Eq.~(\ref{elcrossed2}).
This means interference with maximal contrast, corresponding to the
maximal possible enhancement factor of two. 

\subsubsection{Inelastic contribution}

The inelastic component $I_{\rm in}$ of the intensity, finally, arises from
two-photon scattering. Here, the overlap $\langle g_1|g_1\rangle$, see
Eq.~\ref{i3}, again implies a sum over the undetected
photon, which now may have a frequency different from $\omega_L$.
With two atoms, $|g_1\rangle$ is a sum of four different contributions,
corresponding to the diagrams (2a-d).
Correspondingly, we obtain diagonal terms ($\rm |2a|^2,\dots,|2d|^2$), which
contribute to the background signal, and
interference terms, which may contribute to
the backscattering cone, see below.

Let us examine first the diagrams (2a) and (2b),
where the elastic
scattering event occurs before the inelastic one. Here, the single-atom
scattering amplitude is multiplied by a constant factor $B(\omega_L)$.
This means that - apart from the modification of the prefactor $\eta$ - 
both $\rm |2a|^2$ and $\rm |2b|^2$ give 
the same result as in the single-atom case, Eqs.~(\ref{i3},\ref{in}):
\begin{equation}
I_{\rm II}=\frac{\tilde{\eta}}{\eta}\int d\omega P^{(in)}(\omega)=
{\tilde{\eta}}\frac{s^2}{2}.\label{lin1}
\end{equation}
In the other two cases (2c) and (2d),
the frequency to be inserted
in the factor $B(\omega)$, Eq.~(\ref{beta}), equals the final frequency of
the detected photon
(or - equivalently - of the undetected one, since 
$|B(\omega)|^2=|B(2\omega_L-\omega)|^2$). Hence, a factor $|B(\omega)|^2$ must
be inserted in the integral over the
inelastic power spectrum, Eq.~(\ref{inintegral}).
The resulting integral
can be easily performed, and yields:
\begin{eqnarray}
I_{\rm I} & = & \frac{\tilde{\eta}}{\eta}\int d\omega 
\left|\frac{\omega_L-\omega_0}{\omega-\omega_0}\right|^2
P^{(in)}(\omega)\nonumber\\ 
& = & {\tilde{\eta}} \frac{s^2}{2}
\left(\frac{3}{4}+\frac{\delta^2}{\Gamma^2}\right).\label{lin2}
\end{eqnarray}
Hence, the four diagonal terms $\rm |2a|^2,\dots,|2d|^2$, give the
following contribution to the
inelastic background intensity:
\begin{equation}
L^{\rm (in)}=2I_{\rm I}+2I_{\rm II}=\tilde{\eta} \left(\frac{7}{4}+\frac{\delta^2}{\Gamma^2}\right)
s^2.\label{lin3}
\end{equation}
Note that, for $\delta=0$, the contribution from
(\ref{lin2}) is smaller than the one from (\ref{lin1}) (by a factor $3/4$).
This is due to the
fact that, after the inelastic scattering event, the photon frequencies
are no longer exactly on resonance, see Fig.~\ref{spec}(a), what reduces
the cross section of the scattering by the other atom. The opposite is the
case for large detuning $\delta$: here, the inelastic scattering
brings one of the two photons close to the atomic resonance,
see Fig.~\ref{spec}(b), thereby increasing the corresponding contribution
to the background signal. 

The inelastic component of the enhanced backscattering signal
arises from the interference of
(2a) with (2d), and (2b) with (2c).
(Remember that every diagram 
interferes only with those where the undetected photon is emitted by the
same atom.) As argued above, equality of the corresponding geometrical
phases, and thereby full constructive interference, is guaranteed if
the wavevector of the detected photon is opposite to the incident 
wavevector, i.e., ${\bf k}_D=-{\bf k}_L$. Obviously, this condition will
not be exactly fulfilled in the presence of inelastic scattering, even in exact
backscattering direction (since in general
$|{\bf k}_D|\neq |{\bf k}_L|$). The difference
can be neglected, however, if we assume that the atomic linewidth $\Gamma$
and the detuning $\delta=\omega_L-\omega_{\rm at}$,
i.e. the parameters which determine the width of the
power spectrum, see Fig.~\ref{spec},
are much smaller than
the inverse of the distance $r_{12}$ between the atoms:
\begin{equation}
\delta,\Gamma\ll \frac{c}{r_{12}}\label{dgllr}.
\end{equation}
In other words: the propagation time $r_{12}/c$ between the atoms
is much smaller than the timescales associated with $\delta$ and $\Gamma$.
This condition ensures a vanishing geometric phase difference, i.e.
$\exp[({\bf k}_L+{\bf k}_D)\cdot
({\bf r}_1-{\bf r_2})]\simeq 1$,
and is well fulfilled in the 
experiment \cite{Chan03}. What remains is the integration over the inelastic
spectrum, taking into account the photon exchange factors 
$B(\omega_L)$ or $B(\omega)$ in the cases (2a,b) or (2c,d),
respectively:
\begin{equation}
2 \int d\omega {\rm Re}\left\{\frac{\omega_L-\omega_0}{\omega-\omega_0}
\right\}
P^{\rm (in)}(\omega)=
 \frac{3}{4}\eta s^2.\label{cin}
\end{equation}
Here, we have neglected the exponential factor
$e^{i(\omega-\omega_L)r_{12}}\simeq 1$
describing the propagation in the vacuum - the same approximation as
above, see Eq.~(\ref{dgllr}).
From the two interfering pairs of diagrams, the inelastic contribution
to the backscattering signal is obtained as twice
the result of Eq.~(\ref{cin}) with modified
prefactor $\tilde{\eta}$:
\begin{equation}
C^{(\rm in)} = 
\tilde{\eta} \frac{3}{2}s^2.\label{cin2}
\end{equation}  
Note that $C^{(\rm in)}$ is strictly smaller than the inelastic
background, Eq.~(\ref{lin3}), which leads to
a reduction of
the backscattering enhancement factor, see below. This
is consistent with the fact that
two interfering diagrams, e.g. (2a) and (2d),
are no more linked by the reciprocity symmetry: only diagrams with identical
initial and final photon frequencies
interfere with each other, whereas the reciprocity symmetry connects diagrams
where initial and final frequencies are exchanged.  

\subsubsection{Double scattering enhancement factor}

Adding all contribution, we have:
\begin{eqnarray}
L = L^{(1)}+L^{(2,\rm el)}+L^{(\rm in)} & = & \tilde{\eta} \left(s-\frac{9}{4}s^2+\frac{\delta^2}{\Gamma^2}s^2\right) \\
C = C^{(1)}+C^{(2,\rm el)}+C^{(\rm in)}& = & \tilde{\eta}\left(s-\frac{5}{2}s^2\right).
\end{eqnarray}
Finally, the double scattering enhancement factor reads:
\begin{equation}
\alpha=\frac{L+C}{L} = \frac{8-(19-4\delta^2/\Gamma^2)s}{4-(9-4\delta^2/\Gamma^2)s}\label{enhance}.
\end{equation}
Remember that single scattering has been removed by the
helicity-preserving polarization channel. 

At this stage, it
is convenient to introduce the saturation parameter on resonance:
\begin{equation}
s_0~=~\frac{2 d^2 I_{\rm in}}{\Gamma^2/4}~=
\left(1+\frac{4\delta^2}{\Gamma^2}\right) s,\label{s0}
\end{equation}
which depends only on the incident intensity $I_{\rm in}$ (and not on the
detuning $\delta$). Then, Eq.~(\ref{enhance}) can be rewritten:
\begin{equation}
\alpha~=~\frac{2+x}{1+x},\label{enhance2}
\end{equation}
with
\begin{equation}
x=\frac{s_0}{4-10 s}\simeq \frac{s_0}{4}.\label{x}
\end{equation}
Here, we have used that $s$ is small - otherwise,
our perturbative treatment (two-photon scattering) would be invalid.
If the detuning $\delta$ is of the order of the linewidth 
\begin{figure}
\centerline{\epsfig{figure=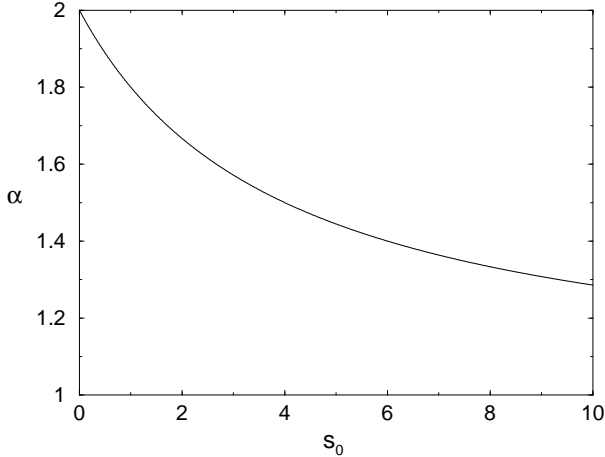,height=8cm,angle=270}}
\caption{The enhancement factor $\alpha=(8+s_0)/(4+s_0)$ as a function of the
incident
intensity $s_0$, and large detuning $\delta=\omega_L-\omega_{\rm at}$.
If $\delta$ is not large, the displayed curve is valid only up to intensities 
$s_0\ll 1+4\delta^2/\Gamma^2$,
corresponding to a small saturation parameter $s\ll 1$, cf. Eq.~(\ref{s0}).
}
\label{alpha}
\end{figure}
$\Gamma$, this implies that $s_0$ must also be small. In this case,
Eq.~(\ref{enhance2}) yields:
\begin{equation}
\alpha\simeq 2-\frac{s_0}{4}.\label{linear}
\end{equation}
In principle, however, we may choose also a large value of the
detuning $\delta$, as long as we stay near-resonant, and fulfill
$(\delta/\Gamma)^2\ll 1/(\omega r_{12})^2$.
\footnote{This
condition implies $s_0|B(\omega_{\rm at})|^2\ll 1$,
see Eqs.~(\ref{beta},\ref{s0}), and thereby
suppresses exchange of more than one resonant photon between
the two atoms, leading to terms proportional 
to $s_0^2$ (or higher order).}
This means that $s_0$ may be large
although $s$ is small, see Eq.~(\ref{s0}). In that case, the enhancement
factor is given by Eq.~(\ref{enhance2}), with $x=s_0/4$, see Fig.~\ref{alpha}.
This equation is valid for all values of $s_0$ corresponding to small $s$,
i.e. $s_0\ll 1+4\delta^2/\Gamma^2$.

It may appear surprising that the enhancement factor $\alpha$ depends 
only on the intensity $s_0$ of the incident light,
see Eqs.~(\ref{enhance2},\ref{x}),
whereas the intensity scattered by a
{\em single} atom is determined by the saturation parameter 
$s$, see Sec.~\ref{spower}.
This result is related
to the form of the inelastic spectrum, see Fig.~\ref{spec}: since one of
the two photons is always close to the atomic resonance after the
inelastic scattering, the asymmetry between the reversed paths 
(see the following section) is larger
for larger initial detuning $\delta$, at a given value of $s$. 
Thereby, we can understand why, at a fixed $s$,
the enhancement factor $\alpha$ decreases when increasing $\delta$.
However, we are not aware of an intuitive explanation why the
relevant parameter turns out to be $s_0$, and not some other,
similar combination of $\delta$ and $s$.

\begin{figure}
\centerline{\epsfig{figure=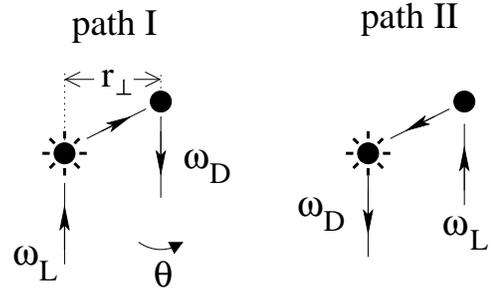,width=7cm,angle=0}}
\caption{Two reversed scattering paths, whose interference gives rise
to enhanced backscattering. The left atom scatters inelastically,
changing the photon frequency from $\omega_L$ to $\omega_D$.
Consequently, the amplitudes
of the elastic scattering event by the right atom are different for both
paths, see Eqs.~(\ref{path1},\ref{path2}). This leads to a reduced contrast of
the interference pattern as a function of the detection angle $\theta$, see
Fig.~\ref{theta}.}
\label{paths}
\end{figure}

\subsection{Interpretation: which-path information and coherence loss}
\label{interpret}

In this section, we discuss the
physical mechanism responsible for the reduction of the
backscattering enhancement factor. As we have seen above,
it originates solely from inelastic scattering. For this reason, we will
only consider inelastic scattering in the
following. 

\subsubsection{Interference between counterpropagating paths}

Generally, coherent backscattering arises from constructive interference
between two scattering paths where the detected photon interacts with the
respective scatterers in opposite order.
The maximum enhancement factor of two
is obtained if every path has a counterpropagating counterpart with
the same amplitude.
In the case of two photons,
a \lq scattering path\rq\ in principle also specifies the final state 
$|{\bf k}\rangle$ of the
undetected photon. As we have seen above, Eq.~(\ref{angint}),
the average over the angular variables of the undetected photon 
destroys interference between paths where the inelastic scattering occurs
at different atoms (if the 
atoms are far away from each other). 
Consequently,
if we concentrate on the
detected photon, we should compare only the two reversed
paths where the inelastic
scattering occurs at the same atom, and
the final frequency $\omega_D=2\omega_L-\omega_{\bf k}$ is the same (due
to energy conservation), as shown in Fig.~\ref{paths}.
Here, the left atom is marked as the one which scatters inelastically.
Neglecting the propagation in the vacuum, see Eq.~(\ref{dgllr}),
the amplitudes $E_{\rm I,II}$ of the two reversed 
paths are obtained by multiplying
the scattering amplitudes of the elastic and inelastic scattering event, 
Eqs.~(\ref{t1},\ref{t2}). 
Since the elastic scattering 
occurs at two different frequencies
$\omega_L$ and $\omega_D$, the amplitudes 
are not identical:
\begin{eqnarray}
E_{\rm I} & = & 
\left(\frac{1}{\omega_D-\omega_0}+
\frac{1}{2\omega_L-\omega_D-\omega_0}\right)
\frac{e^{-ikr_\perp\theta/2}}{\omega_D-\omega_0}\label{path1}\\
E_{\rm II} & = & 
\left(\frac{1}{\omega_D-\omega_0}+
\frac{1}{2\omega_L-\omega_D-\omega_0}\right)
\frac{e^{ikr_\perp\theta/2}}{\omega_L-\omega_0}\label{path2},
\end{eqnarray}
where $\theta$ denotes the angle between detector and backscattering
direction, $r_\perp$ the perpendicular distance between the atoms,
and prefactors not depending on $\omega_D$ or $\theta$ are neglected.

The asymmetry of amplitudes leads to a loss of contrast in the interference
pattern:
\begin{equation}
|E_{\rm I}+E_{\rm II}|^2=|E_{\rm I}|^2+|E_{\rm II}|^2+
2|E_{\rm I}||E_{\rm II}|\cos\left(\phi_0+
kr_\perp\theta\right),\label{e1e2}
\end{equation}
if we plot the intensity as a function of the detection angle $\theta$,
see Fig.~\ref{theta}(a). Here, the value of the phase shift reads:
\begin{equation}
\tan(\phi_0)=\frac{2\left(\delta-\epsilon\right)\Gamma}
{4\delta\epsilon+\Gamma^2},\label{phi0}
\end{equation}
and $-\pi<\phi_0<\pi$, where $\epsilon=\omega_D-\omega_{\rm at}$ is the
detuning of the detected photon.
The maximum contrast,
where the intensity oscillates between zero and two times the mean value
as a function of $\theta$, 
is achieved only in the
case $|E_{\rm I}|=|E_{\rm II}|$. Nevertheless, also for
$|E_{\rm I}|\neq |E_{\rm II}|$, the two amplitudes interfere
perfectly coherently: the contrast is as large as it can be, given that
the probabilities $|E_{\rm I,II}|^2$ of path I and II are different.
This case is analogous to a double-slit experiment performed with
a perfect monochromatic plane wave, but with different slit sizes.
Here also, the contrast is reduced although the wave coherence is perfectly
preserved \cite{BornWolf}.

\subsubsection{Second photon as a which-path detector}

What we have said so far is valid if the final frequency
$\omega_D$ is fixed.
In reality, however, $\omega_D$ is a random variable. This implies
that the ratio $E_{\rm I}/E_{\rm II}$ between the amplitudes of both
paths fluctuates randomly, leading to a loss of coherence between the two
paths.
\footnote{In general, fluctuations of the phase and of
the absolute value of $E_{\rm I}/E_{\rm II}$ both lead to 
$\gamma_{\rm I,II}<1$, as defined in Eq.~(\ref{chi}).
In our case, the phase fluctuations have a stronger impact,
at least for moderate values of the detuning $\delta$ (not much larger than
$\Gamma$).}
If we look at the interference pattern of the average intensity,
integrated over $\omega_D$, the loss of coherence reveals itself
as a reduced contrast $I_{\rm max}-I_{\rm min}<4\sqrt{I_{\rm I}I_{\rm II}}$,
i.e. smaller than
the maximum value for
two perfectly coherent waves of intensities $I_{\rm I}$ and $I_{\rm II}$,
respectively, see Eq.~(\ref{e1e2}). The factor $\gamma_{\rm I,II}$ 
by which the contrast is reduced is called {\em degree of coherence}
(see \cite{BornWolf}, p.~499-503):
\begin{eqnarray}
I & = & \int d\omega_D |E_{\rm I}(\omega_D)+E_{\rm II}(\omega_D)|^2\\
& = & I_{\rm I}+I_{\rm II}+2\sqrt{I_{\rm I}I_{\rm II}}~\gamma_{\rm I,II}~
\cos\left(\phi+kr_\perp\theta\right),\label{chi}
\end{eqnarray}
with $I_{\rm I}$ and $I_{\rm II}$ the average intensities of path I and II,
respectively,
given by Eqs.~(\ref{lin1},\ref{lin2}), and $\phi$ the remaining phase shift.
An example is shown
in Fig.~\ref{theta}(b). 
To make the loss of coherence
visible, we have indicated by the arrows
the contrast obtained for the same values of $I_{\rm I,II}$,
but $\gamma_{\rm I,II}=1$. 

An alternative physical explanation of the coherence loss can be obtained
by an analogy to Young's famous double-slit experiment. As it is well
known, interference is necessarily destroyed whenever we observe
which slit the particle passes through (see, e.g., \cite{pfau94}).
If $|D_{\rm I}\rangle$ and
$|D_{\rm II}\rangle$ denote the quantum states of the which-path detector
corresponding to path I and II, 
the degree of coherence is obtained as the overlap of the normalized detector
states \cite{Tan93}:
\begin{equation}
\gamma_{\rm I,II}=\frac{\left|\langle D_{\rm I}|D_{\rm II}\rangle\right|}
{\sqrt{\langle D_{\rm I}|D_{\rm I}\rangle\langle D_{\rm II}|
D_{\rm II}\rangle}},
\label{chi2}
\end{equation}
This implies perfect coherence, $\gamma_{\rm I,II}=1$, if the paths are
indistinguishable
(i.e. if the detector states are identical),
and total loss of coherence, $\gamma_{\rm I,II}=0$,
if the paths can be distinguished with
certainty (i.e. if the detector states are orthogonal).

In our case, the path detector is given by the undetected photon.
Remember that its frequency  is correlated 
to the one of the detected photon,
due to conservation of energy at the inelastic scattering event.
Therefore, the different dependence of the amplitudes $E_{\rm I,II}$
of path I and II
on the
frequency of the detected photon, see Eqs.~(\ref{path1},\ref{path2}),
reflects itself in the final state of the undetected
photon:
\begin{eqnarray}
|D_{\rm I}\rangle & = & \sum_{{\bf k}\epsilon} E_{\rm I}(2\omega_L-
\omega_{\bf k})
(\epsilon_L\cdot\epsilon^*)|{\bf k}\epsilon\rangle,\label{d1}\\
|D_{\rm II}\rangle & = & \sum_{{\bf k}\epsilon} E_{\rm II}(2\omega_L-
\omega_{\bf k})
(\epsilon_L\cdot\epsilon^*)|{\bf k}\epsilon\rangle\label{d2}.
\end{eqnarray}
Since $|D_{\rm I}\rangle$ and
$|D_{\rm II}\rangle$ are not identical,
the state
of the undetected photon contains information about which path 
the first photon has taken. This leads to a loss
of coherence according to Eq.~(\ref{chi2}).
Using Eqs.~(\ref{path1},\ref{path2}), we obtain:
\begin{eqnarray}
\gamma_{\rm I,II} & = & \sqrt{\frac{9+4\delta^2/\Gamma^2}{12+16\delta^2/\Gamma^2}},\\
\tan(\phi) & = & \frac{2\delta}{3\Gamma},\label{phi}
\end{eqnarray}
and $-\pi<\phi<\pi$. The probabilities of the two paths,
$I_{\rm I}=\langle D_{\rm I}|D_{\rm I}\rangle$ and
$I_{\rm II}=\langle D_{\rm II}|D_{\rm II}\rangle$, respectively,
have been calculated in the previous section, see Eqs.~(\ref{lin1},\ref{lin2}).
Note that the phase shift, Eq.~(\ref{phi}),
vanishes for zero detuning - this can be traced back
to the symmetry of the power spectrum with respect to
$\omega_{\rm at}$, see Fig.~\ref{spec}(a), and the fact that the
scattering amplitude 
$(\omega_L-\omega_0)^{-1}=-2i/\Gamma$ is purely
imaginary at resonance, $\omega_L=\omega_{\rm at}$.

One may wonder how the which-path
information can be extracted from the second photon.
In principle, this is possible,
e.g., by performing a projection $|D_{\rm I}\rangle\langle D_{\rm I}|$,
with possible measurement results \lq 0\rq\ and \lq 1\rq.
If we measure \lq 0\rq,
we know with certainty that path II has been taken, whereas
in the case \lq 1\rq, both paths are possible, but with increased
probability of path I.
Due to the rather complicated expressions of
$|D_{\rm I}\rangle$ and $|D_{\rm II}\rangle$, see Eqs.~(\ref{d1},\ref{d2}),
it seems rather difficult to build such a detector in practice.
The simplest observation is to look at the frequency 
$\omega_{\bf k}=2\omega_L-\omega_D$.
Since
$|E_{\rm I}(\omega_D)|^2$ and $|E_{\rm II}(\omega_D)|^2$ depend differently on
$\omega_D$,
this observation does
give some information about the path, although it does not
resolve the phase dependence of $E_{\rm I}(\omega_D)$ and
$E_{\rm II}(\omega_D)$.
Nevertheless, the state of the second photon
is completely determined by $\omega_{\bf k}$ (since the angular 
distribution is given by the polarization, independent of I or II).
This implies that the which-path information can be erased
by putting a spectral filter in front of the detector. Then,
the measurement of the frequency gives no information (since it is
always the same, determined by the filter), and, consequently, the
coherence is fully restored, see Fig.~\ref{theta}(a).

\subsubsection{Coherence between the light emitted by atom 1 and 2}

At first sight, it may seem surprising that, for $\delta>0$,
the maximum of the interference
pattern is
not found in the backscattering direction, at $\theta=0$. The reason is that
we have specified the
inelastically scattering atom, see Fig.~\ref{paths}, thereby introducing
an asymmetry.
However, with equal probability, the inelastic scattering may occur 
at  the other atom, and then the interference pattern is shifted in the
opposite direction. As already mentioned above, these two cases do not
interfere, since they are distinguished by the undetected photon
(in a similar way as the coherence between path I and II is reduced).
If we add therefore both interference patterns incoherently,
the new maximum is found at $\theta=0$ (as it should be), and the contrast
of the interference pattern will be further reduced, by a factor $\cos(\phi)$:
\begin{equation}
I_{\rm tot}=\underbrace{2I_{\rm I}+2I_{\rm II}}_{\displaystyle
L^{(\rm in)}}+\underbrace{4\sqrt{I_{\rm I}I_{\rm II}}~\gamma_{\rm I,II}
\cos(\phi)}_{\displaystyle C^{(\rm in)}}~\cos(kr_\perp\theta).
\label{itot1}
\end{equation}
In the backscattering direction, $\theta=0$, we recover the  total intensity
as a sum of a background and an enhanced backscattering term
$L^{(\rm in)}$ and $C^{(\rm in)}$, which we have calculated in the
previous section, see Eqs.~(\ref{lin3},\ref{cin2}).

What degree of coherence do we associate to the total interference pattern?
Note
that we are now dealing with
four different diagrams, see Fig.~\ref{2atom2},
whereas (second-order) coherence is a property of
{\em two} interfering waves. Therefore, before we speak of coherence,
we have to specify how to
divide the
four diagrams into two waves. For that purpose,
we consider the light emitted by atom 1 as the first wave, see
diagrams (2a) and (2c),
and the light by atom 2 as the second one, see diagrams (2b) and (2d).
Indeed let us point out, even if it is obvious, that the
total scattered electrical field is the sum of the electrical fields
radiated by each atom in response to the {\em local}
electrical field. This local field embodies the incident field {\em and} 
the field radiated by the other atom. This means that multiple scattering
is taken into account.
Thus, the total radiated intensity contains
interference terms between the fields radiated by each atom, some of them
surviving the spatial and spectral averaging in the backscattering direction.
In this respect, the CBS signal probes the {\em spatial coherence} between 
the two radiated fields.

For reasons of symmetry, the intensities emitted by atom 1 and 2 are identical:
\begin{equation}
I_1=I_2=I_{\rm I}+I_{\rm II}.\label{ii}
\end{equation}
According to the definition of the degree of coherence,
\begin{equation}
I_{\rm tot}=I_1+I_2+2\sqrt{I_1I_2}~\gamma_{1,2}~\cos(kr_\perp\theta),\label{itot2}
\end{equation} 
see Eq.~(\ref{chi}),
and taking into account Eqs.~(\ref{itot1},\ref{ii}), we obtain
\begin{equation}
\gamma_{1,2}=\gamma_{\rm I,II}\cos(\phi)\frac{2\sqrt{I_{\rm I}I_{\rm II}}}
{I_{\rm I}+I_{\rm II}}.\label{chi12}
\end{equation}
On the other hand, if we compare (\ref{chi12}) with the underbraced
terms in Eq.~(\ref{itot1}), the close relation between
$\gamma_{1,2}$ and the 
backscattering enhancement factor 
$\alpha^{(\rm in)}$ (considering only inelastic scattering)
becomes evident:
\begin{eqnarray}
\gamma_{1,2} & = & 
\frac{C^{(\rm in)}}{L^{(\rm in)}} = \alpha^{(\rm in)}-1\label{gammaalpha}\\
& = & \frac{6}{7+
4\delta^2/\Gamma^2}.
\end{eqnarray} 
Here, we have inserted the results (\ref{lin3},\ref{cin2}) of the
previous section. From Eq.~(\ref{chi12}), we see that the coherence
$\gamma_{1,2}$ between atom 1 and 2 is reduced by the average
over the power spectrum, leading to $\gamma_{\rm I,II}<1$, on the one hand,
and by the random choice of atom 1 or 2 as the
inelastically scattering atom, on the other one. The latter affects the
coherence both due to different phases $\pm\phi$ and different amplitudes
$I_{\rm I}$ and $I_{\rm II}$. For not too large values
of the detuning $\delta$, the amplitude term can almost be neglected 
(being larger than $96\%$ for $\delta<\Gamma$), whereas the phase term
$\cos(\phi)$, although strictly vanishing at $\delta=0$, gives a significant
contribution if $\delta$ is of the order of $\Gamma$, see Eq.~(\ref{phi}).

Let us note that the above interpretation in terms of a which-path
experiment remains valid when adding the new pair of diagrams.
Then, the detector states $|D_{\rm 1,2}\rangle$ are given by the state of the
undetected photon represented by diagram $(2a)+(2c)$,
on the one hand, and $(2b)+(2d)$, Fig.~\ref{2atom2},
on the other one. Similarly, we may also
include the elastic component of the photodetection 
\begin{figure}
\centerline{\epsfig{figure=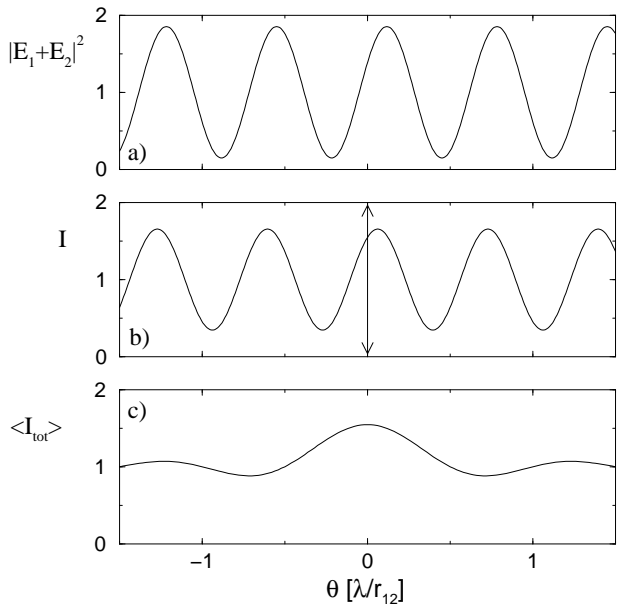,height=8cm,angle=270}}
\caption{Intensity of the inelastic detection signal (suitably normalized),
as a function of the
angle $\theta$ between detector and incident light, for detuning
$\delta=\Gamma$. In (a), both the final
frequency $\omega_D$ and the positions ${\bf r}_{12}$ of the elastically and
inelastically scattering atom are fixed ($\omega_D=\omega_{\rm at}$ and
$r_\perp=2r_{12}/3$), (b) shows the average over
$\omega_D$, and (c) the average over both $\omega_D$ and ${\bf r}_{12}$.
To make the loss of coherence due to the average over the power spectrum
visible, the arrows in (b) indicate the contrast
for perfect coherence, $\gamma_{\rm I,II}=1$, but same asymmetry $I_{\rm I}/I_{\rm II}$,
cf. Eq.~(\ref{chi}). The average over the positions, apart from
washing out the side maxima,
leads to
a further loss of coherence, due to exchange of
the inelastically and elastically scattering atom (see main text).}
\label{theta}
\end{figure}
signal, where the
undetected photon is described by the diagrams
in Fig.~\ref{2atom1}. Furthermore, also the relation between degree of
coherence, $\gamma_{12}$, and enhancement
factor $\alpha$, Eq.~(\ref{gammaalpha}), remains valid for the total signal
(as it is always the case if $I_1=I_2$):
\begin{equation}
\gamma_{12}^{(\rm el+in)} =
\frac{C}{L} = 
\alpha-1 = \frac{4}{4+s_0},
\end{equation}
where we have inserted Eqs.~(\ref{enhance2},\ref{x}) of the previous section.

The average over the positions ${\bf r}_1$ and ${\bf r}_2$ of the atoms,
finally, does not affect the intensity observed at $\theta=0$;
it only reduces the side maxima, and determines the shape of the
backscattering \lq cone\rq. As an example, Fig.~\ref{theta}(c) shows the
result for an angular average over ${\bf r}_2-{\bf r}_1$, at fixed
distance $r_{12}$, where the cone shape is described by the function
$\sin(x)/x$. 

Finally, we want to stress that there is no
loss of coherence associated with the inelastic
scattering \lq on its own\rq, but only in connection with
the frequency filtering induced by the elastic scattering event.
This can be demonstrated as follows: 
let us imagine that the response of the second atom
is frequency-independent, i.e. $B(\omega)={\rm const}$ in
Eq.~(\ref{beta}). Then, the amplitudes of two reversed paths,
see Eqs.~(\ref{path1},\ref{path2}), are identical, 
the undetected photon does not carry any which-path
information, and we recover 
the enhancement factor two, even in the presence
of inelastic scattering.
Such a situation can be realized, e.g.,
by choosing atoms with different linewidths $\Gamma_2\gg\Gamma_1$,
such that atom 2 cannot resolve the spectrum emitted by atom 1.
In this case, a significant reduction of the enhancement factor is observed
only if we increase the distance $r_{12}$ between the atoms,
such that the propagation in the vacuum becomes relevant.  
 
\subsection{Conclusion}
\label{concl}

In summary, we have presented the first calculation of coherent 
backscattering in the presence of saturation.
For two distant atoms, with single scattering excluded,
the slope of the backscattering
enhancement factor as a function of
the incident intensity $s_0$ at $s_0=0$ equals $-1/4$, independently of the
value of the detuning.
The reduction of the enhancement factor can be traced
back to the following two random processes:
firstly, the
frequency of the photons may be changed by the inelastic scattering event,
which may, secondly, occur either at the first or at the second atom.
Both processes (the latter one only for nonzero detuning, see Eq.~(\ref{phi}))
lead to a random phase shift between the doubly scattered
light emitted by the first atom, on the one hand, and by the
second atom, on the other one, resulting in a loss of coherence.
Alternatively, the coherence loss can be explained by
regarding the undetected photon as a which-path detector:
its final state contains
information about whether the detected photon has been
emitted by the first or second atom, thereby partially destroying
coherence between those paths.

Starting from the solution of our model, we can think of
extending it to more general scenarios in two different directions,
either increasing the number of photons, to reach higher 
values of the saturation parameter, or the number of atoms, to treat a
disordered medium of atoms. Since the complexity of the scattering
approach increases dramatically with the number of scattered particles,
it may be more promising to use other methods, such as the
optical Bloch equations \cite{CCT}, in the case of high saturation.
The opposite is true for a large number of scatterers, where we can resort
to known concepts from the theory of multiple scattering. An important
question, which must be solved in order to interpret the results of the
experiment \cite{Chan03}, is how the average propagation of the two-photon
state in the atomic medium affects the coherent backscattering signal.

\acknowledgements

It is a pleasure to thank Cord M\"uller, Vyacheslav Shatokhin,
David Wilkowski, Guillaume Labeyrie, and Andreas Buchleitner
for fruitful discussions, critical remarks, and interest in our work.
T.W. is indebted to the Deutsche Forschungsgemeinschaft for
financial support within the Emmy Noether program.
Laboratoire Kastler Brossel is laboratoire 
de l'Universit{\'e} Pierre et Marie
Curie et de l'Ecole Normale Sup{\'e}rieure, UMR 8552 du CNRS. CPU time on 
various computers has been provided by IDRIS.

\appendix

\section{Two-atom scattering matrix}
\label{app2at}
\begin{figure}
\centerline{\epsfig{figure=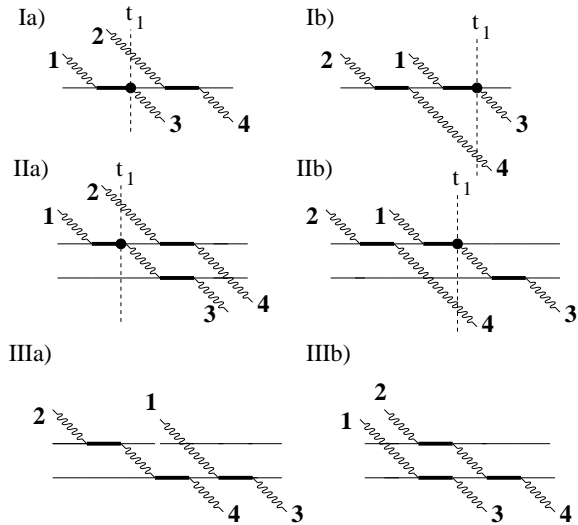,width=8cm,angle=0}}
\caption{Diagrams describing scattering of two photons by a single atom 
(Ia,b) and two atoms, (IIa,b) and (IIIa,b). The curly lines represent photons,
and thin or thick lines an atom in the ground or excited state.
In order to simplify the
comparison between (I) and (II), we split the
diagrams into a right and a left half (see main text).}
\label{doublefig}
\end{figure}

In this appendix, we calculate the scattering of two photons by two atoms.
For this purpose, we use the following expansion of the evolution
operator $U(t_0,t)=\exp[-i(H_0+V)(t-t_0)]$:
\begin{eqnarray}
U(t_0,t) & = & \sum_{n=0}^\infty \int_{t_0}^t dt_1\int_{t_1}^t dt_2\dots
\int_{t_{n-1}}^t dt_n\nonumber\\
& &  U_0(t_0,t_1)VU_0(t_1,t_2)V\dots V U_0(t_n,t),\label{u}
\end{eqnarray}
where $U_0(t_0,t)=\exp[-iH_0(t-t_0)]$ denotes the free evolution.
With each interaction $V$, see Eq.~(\ref{v}), an atom may emit a photon,
or absorb one of the two photons. The corresponding \lq paths\rq\ connecting
the initial and final two-photon state 
$|i\rangle=|{\bf k}_1\epsilon_1,{\bf k}_2\epsilon_2\rangle$ and
$|f\rangle=|{\bf k}_3\epsilon_3,{\bf k}_4\epsilon_4\rangle$ can be
represented diagrammatically, see Fig.~\ref{doublefig}.
Here, (Ia,b) describes the scattering of two photons by a single atom
\cite{Dal83},
and (IIa,b) and (IIIa,b) the scattering by two atoms.
Let us first concentrate on (Ia,b) and (IIa,b), where the inelastic scattering
event occurs before the elastic one. 
Note that in (IIa), we have not
specified the order in which the photons are emitted or absorbed.
What we mean by this is a sum over all possible orderings, as indicated in
Fig.~\ref{ind}. As we will see below, however, the sum need not be explicitly 
evaluated.
\begin{figure}
\centerline{\epsfig{figure=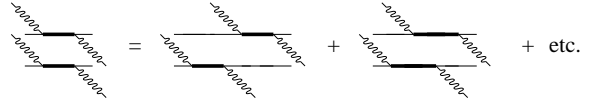,width=8cm,angle=0}}
\caption{Independent scattering of two photon by two different atoms.
This diagram appears as a building block in (IIa) and (IIIb),
Fig.~\ref{doublefig}.}
\label{ind}
\end{figure}

\begin{figure}
\centerline{\epsfig{figure=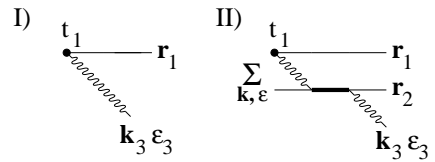,width=6cm,angle=0}}
\caption{(I) Emission of a photon $|{\bf k}_3\epsilon_3\rangle$ at time $t_1$.
(II) Photon emission
and subsequent scattering by the second atom.}
\label{singlefig}
\end{figure}

Furthermore, we have selected one of the
interaction operators $V$ in the expansion (\ref{u}), at which we split the
diagrams
into a right and left half, denoted by $U^{(l,r)}$ in the following.
According to Eq.~(\ref{u}), we may write: 
\begin{equation}
U_{\rm IIa}(t_0,t)~=~\int_{t_0}^t dt_1 U^{(l)}_{\rm IIa}(t_0,t_1)V
U^{(r)}_{\rm IIa}(t_1,t),
\end{equation}
and similarly for the other three diagrams (Ia), (Ib), and (IIb). 
Obviously, the left half is identical in the
one- and two-atom case I and II, respectively. 
In the right half, on the other hand,
the two photons are always independent from each other,
being scattered by different atoms (if at all). This means that the
evolution operator is the product of the two single-photon 
evolution operators:
\begin{equation}
U_{\rm IIa}^{(r)}(t_1,t)~=~U_{\rm II}^{(r,1)}(t_1,t)
U^{(r,2)}_{\rm a}(t_1,t),
\end{equation}
and likewise for (Ia), (Ib), and (IIb). Note that the evolution of the 
first photon ($1\to 3$) depends only on (I) or (II), and not on (a) or (b), 
and vice versa for the second photon.
Thereby, if we want to compare the one- and two-atom case, 
we have to consider only the two single-photon diagrams
$U^{(r,1)}_{\rm I,II}$, which are illustrated
in Fig.~\ref{singlefig}.
The first one simply
describes the emission of photon $|{\bf k}_3\epsilon_3\rangle$
by an atom located at ${\bf r}_1$, followed by free evolution: 
\begin{equation}
VU^{(r,1)}_{\rm I}(t_1,t)
~=~-ig(\epsilon_1\cdot\epsilon_3^*) e^{-i{\bf k}_3\cdot {\bf r}_1} e^{-i\omega_3(t-t_1)}\label{ir1}
\end{equation}
In the second case, the photon is scattered by the other atom. Here, we have
to take the sum over its intermediate state. For the calculation,
it is convenient to
express the time evolution in terms of the corresponding Green's
function:
\begin{equation}
VU^{(r,1)}_{\rm II}(t_1,t)~=~\int_{C^+}\frac{dz}{2\pi i}e^{-iz(t-t_1)}
G^{(r,1)}_{\rm II}(z),\label{ug}
\end{equation}
where the contour $C^+$ runs just above the real axis, i.e., $z=x+i\epsilon$,
$\epsilon>0$, from $x=+\infty$ to $-\infty$, and the Green's function of the
above diagram (II) reads:
\begin{equation}
G^{(r,1)}_{\rm II}(z)  =   \sum_{{\bf k},\epsilon}
\frac{-ig^3(\epsilon_1\cdot\epsilon^*)(\epsilon\cdot\epsilon_3^*)
e^{-i{\bf k}\cdot {\bf r}_1+i({\bf k}-{\bf k}_3)\cdot {\bf r}_2}}{(z-\omega_k)
(z-\omega_0)(z-\omega_3)}.\label{g}
\end{equation}
In the continuous limit ($L\to\infty$), the sum is replaced by an integral
($\sum_k=(L/2\pi)^3 \int dk$). The result of the integral (\ref{g}),
in leading order of $1/(\omega_3 r_{12})$, reads:
\begin{equation}
G^{(r,1)}_{\rm II}(z)  = \frac{3i\Gamma g 
(\epsilon_1\cdot\Delta_{12}\cdot\epsilon_3^*)e^{-i{\bf k}_3\cdot
{\bf r}_2} e^{iz r_{12}}}{4\omega_3 r_{12}(z-\omega_0)(z-\omega_3)}.
\end{equation}
Here, $\Delta_{12}$ denotes the projection onto the plane
orthogonal to ${\bf r}_2-{\bf r}_1$.
Finally, in the contour integral (\ref{ug}), only the pole at $z=\omega_3$
contributes (if $t-t_1\gg 1/\Gamma$): 
\begin{eqnarray}
VU^{(r,1)}_{\rm II}(t_1,t) & = & 
i g(\epsilon_1\cdot\Delta_{12}\cdot\epsilon_3^*)
e^{-i{\bf k}_3\cdot{\bf r}_2}e^{-i\omega_3(t-t_1)}\nonumber\\
& & \times \frac{3\Gamma e^{i\omega_3 r_{12}}}
{4\omega_3 r_{12}(\omega_3-\omega_0)}.\label{iir1}
\end{eqnarray}
Comparing Eqs.~(\ref{ir1},\ref{iir1}), we see that the contribution to the
two-atom scattering matrix represented by (IIa,b)
is given by the
one-atom matrix $S_{\rm I}$,
times a correction of the geometrical phase and the polarization, 
times the photon exchange
factor $B(\omega_3)$, see Eq.~(\ref{beta}).
\begin{eqnarray}
\langle {\bf k}_3\epsilon_3,{\bf k}_4\epsilon_4|S_{\rm II}|
{\bf k}_1\epsilon_1,{\bf k}_2\epsilon_2\rangle & = &
\langle {\bf k}_3\epsilon_3,{\bf k}_4\epsilon_4|S_{\rm I}|
{\bf k}_1\epsilon_1,{\bf k}_2\epsilon_2\rangle\nonumber\\
& & \!\!\!\!\!\!\!\!\!\!\!\!\!\!\!\!\!\!\!\!\!\!\!\!\!\!\!\!\!\!\!\!\times
e^{i{\bf k}_3\cdot({\bf r}_1-{\bf r}_2)}
\frac{(\epsilon_1\cdot\Delta_{12}\cdot\epsilon_3^*)}
{(\epsilon_1\cdot\epsilon_3^*)}
B(\omega_3).\label{a1}
\end{eqnarray}
What remains is the contribution, where the elastic scattering
occurs before the inelastic one, represented by diagrams (IIIa,b) 
in Fig.~\ref{doublefig}.
The calculation can be repeated in almost the same way as above,
or simply by noting that (IIIa,b) is related to (IIa,b) through time reversal,
and the result is:
\begin{eqnarray}
\langle {\bf k}_3\epsilon_3,{\bf k}_4\epsilon_4|S_{\rm III}|
{\bf k}_1\epsilon_1,{\bf k}_2\epsilon_2\rangle & = &
\langle {\bf k}_3\epsilon_3,{\bf k}_4\epsilon_4|S_{\rm I}|
{\bf k}_1\epsilon_1,{\bf k}_2\epsilon_2\rangle\nonumber\\
& & \!\!\!\!\!\!\!\!\!\!\!\!\!\!\!\!\!\!\!\!\!\!\!\!\!\!\!\!\!\!\!\!\times
e^{-i{\bf k}_1\cdot({\bf r}_1-{\bf r}_2)}
\frac{(\epsilon_1\cdot\Delta_{12}\cdot\epsilon_3^*)}
{(\epsilon_1\cdot\epsilon_3^*)}
B(\omega_1).\label{a2}
\end{eqnarray}

Here, the photon exchange factor $B(\omega)$ is evaluated at the frequency
of the initial photon. The total scattering matrix is now readily obtained
by adding $S_{\rm II}$ and $S_{\rm III}$, and also including the diagrams
where the two atoms and/or the two photons are exchanged.

\section{The scalar case}
\label{scalar}

In this appendix, we calculate the photodetection signal for
scalar photons. Although they are not suited for coherent backscattering,
since single scattering cannot be excluded, the solution will be useful
for a future comparison with the results obtained from the
optical Bloch equations, which can be solved much more easily in
the scalar case.

As in the vectorial case, we consider contributions to the detection signal
up to second order in $1/(\omega_Lr_{12})$. We neglect those
terms whose order in $1/(\omega_Lr_{12})$ is changed by the
angular average over ${\bf r}_{12}$.\footnote{These terms give the
corrections of the average photon propagation induced by 
a disordered medium consisting of only a single atom. 
In the case of many
disordered atoms,
they are taken into account by renormalizing the single-photon 
propagation, in order to describe the mean free path and refractive index
of the atomic medium.} 
 Furthermore, we consider only
contributions which do not oscillate rapidly as a function of $r_{12}$,
i.e. which survive an average over $r_{12}$ over one wavelength.

Firstly, since the two atoms may scatter independently from each other,
we obtain two times the single-atom result, see Eqs.~(\ref{el},\ref{in}):
\begin{eqnarray}
L^{({\rm el},0)} & = & \eta_s(s-2s^2),\label{l0s}\\
L^{({\rm in},0)} & = & \eta_s s^2,
\end{eqnarray}
which contributes to the background intensity $L$.
Here we have to take into account that the lifetime
$\Gamma$, and the prefactor $\eta$, are different in the scalar
and vectorial case, respectively.  Instead of Eqs.~(\ref{gamma},\ref{eta}),
the following expressions hold for scalar photons:
\begin{eqnarray}
\Gamma_s & = & \frac{d^2\omega_{\rm at}^3}{2\pi\epsilon_0}\\
\eta_s & = & \left(\frac{\Gamma_s}{2d\omega_L R}\right)^2.
\end{eqnarray}

Next, we consider the cases where one photon is exchanged between the
two atoms. These contribute to the detection signal in second order of
$1/(\omega_Lr_{12})$.
Concerning one-photon scattering,
only the diagrams (1d,e), Fig.~\ref{2atom1}, are relevant,
and we obtain the same result as for  
the $h\parallel h$ channel in the vectorial case, see Eq.~(\ref{elladder1}):
\begin{equation}
L^{({\rm el},1)}=C^{({\rm el},1)}=\eta_s |B|^2 s,
\end{equation}
but with modified \lq photon exchange factor\rq
\begin{equation}
B=\frac{\Gamma}{2\omega_L r_{12}(\omega_L-\omega_0)},
\end{equation}
compare Eq.~(\ref{beta}).

The elastic contribution
quadratic in $s$ arises from interference of two-photon
and one-photon scattering. Let us first look at diagram (2a).
As before in the $h\parallel h$ channel,
it interferes with  (1a+1b)
for the undetected photon, and (1d) or (1e)
for the detected photon,
giving rise to $-\eta_s|B|^2 s^2$ in background $L$ and the cone $C$,
respectively.
Including single scattering, we obtain a new contribution: the detected
photon may be singly scattered (1b), and the undetected
photon either doubly
scattered (1e), or singly-scattered by the other atom (1c).
Here, the state (1e+1c),
of the undetected photon exactly corresponds to
the state (1a+1b),
in the previous case. Consequently, we obtain
another term $-\eta_s|B|^2 s^2$ in the background.

With diagram (2b),
the above considerations can be repeated in almost the
same way. The difference from (2a)
is only that the detected photon propagates
in the opposite direction. Consequently, we obtain a new term $-\eta_s|B|^2
s^2$ in the
cone $C$, instead of the background $L$.

Diagram (2e) is identical to diagram (2a),
since we cannot distinguish between singly or doubly scattered photons
(open or full arrows in Figs.~\ref{2atom1}-\ref{2atom3})
in the scalar case. 
(2c), (2d), and (2f), finally, are obtained by exchanging the atoms.
Adding all contributions mentioned above, we get:
\begin{eqnarray}
L^{(\rm el,2)}& = & -10 \eta_s |B|^2 s^2\label{lel2}\\
C^{(\rm el,2)} & = & -8 \eta_s |B|^2 s^2\label{cel2}.
\end{eqnarray}

As for the inelastic component, we only have to include the
new diagrams (2e,f), which - as already mentioned
above - are identical to (2a,b).
Hence, the background
contribution $2I_{\rm I}$, see Eq.~(\ref{lin2}),
is multiplied by a factor $4$, and
the backscattering cone, Eq.~(\ref{cin2}), by a factor $2$. We obtain:
\begin{eqnarray}
L^{\rm (in,2)}& = & 8I_{\rm I}+2I_{\rm II}=\left(\frac{19}{4}+\frac{\delta^2}{\Gamma^2}
\right)~\eta_s|B|^2s^2\\
C^{\rm (in,2)} & = & 3\eta_s |B|^2 s^2.\label{cins}
\end{eqnarray}
What we have not taken into account so far is
interference between
two diagrams where
the undetected photon is emitted by different atoms. According to
Eq.~(\ref{angint}), the angular integral over the undetected photon then yields
$\sin(\omega r_{12})/(\omega r_{12})$. Hence, if one of the two diagrams
contains a photon exchange, we obtain a contribution proportional to $|B|^2$.
However, it can be shown that these contributions are exactly canceled
by other contributions originating from the diagrams (2g,h),
which also have been neglected so far. For example,
the interference of (2g) with
(1c) for the detected photon and (1c+1e)
for the undetected one is canceled by the interference of (2j) with
(1c) for the detected photon and (1e)
for the undetected one. Similarly, the term $\rm |2g|^2$ is canceled
by the interference of (2g) with (2j).
The underlying reason for all these cancellations
is that what the
undetected photon does after the inelastic scattering is irrelevant.
We are only
interested in its norm,
which is not changed by subsequent scattering events
(due to energy conservation). Hence, the final result 
is given by Eqs.~(\ref{l0s}-\ref{cins}).

\end{document}